\documentclass[twocolumn,twocolappendix]{aastex631}

\usepackage{graphicx}   
\usepackage{amsmath}    
\usepackage{multirow}

\usepackage{xcolor}

\shorttitle{DESI LRG Target Selection}
\shortauthors{Zhou et al.}

\graphicspath{{./}{figures/}}

\begin{document}

\title{Target Selection and Validation of DESI Luminous Red Galaxies}

\correspondingauthor{Rongpu Zhou}
\email{rongpuzhou@lbl.gov}

\author[0000-0001-5381-4372]{Rongpu Zhou}
\affiliation{Lawrence Berkeley National Laboratory, 1 Cyclotron Road, Berkeley, CA 94720, USA}

\author[0000-0002-5665-7912]{Biprateep~Dey}
\affiliation{Department of Physics and Astronomy and PITT PACC, University of Pittsburgh, Pittsburgh, PA 15260, USA}

\author[0000-0001-8684-2222]{Jeffrey A.~Newman}
\affiliation{Department of Physics and Astronomy and PITT PACC, University of Pittsburgh, Pittsburgh, PA 15260, USA}

\author[0000-0002-2929-3121]{Daniel~J.~Eisenstein}
\affiliation{Center for Astrophysics $|$ Harvard \& Smithsonian, 60 Garden Street, Cambridge, MA 02138, USA}

\author{K.~Dawson}
\affiliation{Department of Physics and Astronomy, The University of Utah, 115 South 1400 East, Salt Lake City, UT 84112, USA}

\author{S.~Bailey}
\affiliation{Lawrence Berkeley National Laboratory, 1 Cyclotron Road, Berkeley, CA 94720, USA}

\author{A.~Berti}
\affiliation{Department of Physics and Astronomy, The University of Utah, 115 South 1400 East, Salt Lake City, UT 84112, USA}

\author{J.~Guy}
\affiliation{Lawrence Berkeley National Laboratory, 1 Cyclotron Road, Berkeley, CA 94720, USA}

\author{Ting-Wen Lan}
\affiliation{Graduate Institute of Astrophysics and Department of Physics, National Taiwan University, No. 1, Sec. 4, Roosevelt Rd., Taipei 10617, Taiwan}

\author[0000-0002-6684-3997]{H.~Zou}
\affiliation{National Astronomical Observatories, Chinese Academy of Sciences, A20 Datun Rd., Chaoyang District, Beijing, 100101, P.R. China}

\author{J.~Aguilar}
\affiliation{Lawrence Berkeley National Laboratory, 1 Cyclotron Road, Berkeley, CA 94720, USA}

\author{S.~Ahlen}
\affiliation{Physics Dept., Boston University, 590 Commonwealth Avenue, Boston, MA 02215, USA}

\author[0000-0002-3757-6359]{Shadab Alam}
\affiliation{Institute for Astronomy, University of Edinburgh, Royal Observatory, Blackford Hill, Edinburgh EH9 3HJ, UK}

\author{D.~Brooks}
\affiliation{Department of Physics \& Astronomy, University College London, Gower Street, London, WC1E 6BT, UK}

\author{A.~de la Macorra}
\affiliation{Instituto de F\'{\i}sica, Universidad Nacional Aut\'{o}noma de M\'{e}xico,  Cd. de M\'{e}xico  C.P. 04510,  M\'{e}xico}

\author{A.~Dey}
\affiliation{NSF's National Optical-Infrared Astronomy Research Laboratory, 950 N. Cherry Avenue, Tucson, AZ 85719, USA}

\author{G.~Dhungana}
\affiliation{Department of Physics, Southern Methodist University, 3215 Daniel Avenue, Dallas, TX 75275, USA}

\author{K.~Fanning}
\affiliation{Department of Physics, The Ohio State University, 191 West Woodruff Avenue, Columbus, OH 43210, USA}
\affiliation{Center for Cosmology and AstroParticle Physics, The Ohio State University, 191 West Woodruff Avenue, Columbus, OH 43210, USA}

\author{A.~Font-Ribera}
\affiliation{Institut de F\'{i}sica d’Altes Energies (IFAE), The Barcelona Institute of Science and Technology, Campus UAB, 08193 Bellaterra Barcelona, Spain}

\author[0000-0003-3142-233X]{S.~Gontcho A Gontcho}
\affiliation{Lawrence Berkeley National Laboratory, 1 Cyclotron Road, Berkeley, CA 94720, USA}
\affiliation{Department of Physics and Astronomy, University of Rochester, 500 Joseph C. Wilson Boulevard, Rochester, NY 14627, USA}

\author{K.~Honscheid}
\affiliation{Center for Cosmology and AstroParticle Physics, The Ohio State University, 191 West Woodruff Avenue, Columbus, OH 43210, USA}
\affiliation{Department of Physics, The Ohio State University, 191 West Woodruff Avenue, Columbus, OH 43210, USA}

\author[0000-0002-6024-466X]{Mustapha Ishak}
\affiliation{Department of Physics, The University of Texas at Dallas, Richardson, TX, 75080, USA}

\author{T.~Kisner}
\affiliation{Lawrence Berkeley National Laboratory, 1 Cyclotron Road, Berkeley, CA 94720, USA}

\author[0000-0002-5825-579X]{A.~Kov\'acs}
\affiliation{Departamento de Astrof\'{\i}sica, Universidad de La Laguna (ULL), E-38206, La Laguna, Tenerife, Spain}
\affiliation{Instituto de Astrof\'{i}sica de Canarias, C/ Vía L\'{a}ctea, s/n, 38205 San Crist\'{o}bal de La Laguna, Santa Cruz de Tenerife, Spain}

\author[0000-0001-6356-7424]{A.~Kremin}
\affiliation{Lawrence Berkeley National Laboratory, 1 Cyclotron Road, Berkeley, CA 94720, USA}

\author{M.~Landriau}
\affiliation{Lawrence Berkeley National Laboratory, 1 Cyclotron Road, Berkeley, CA 94720, USA}

\author[0000-0003-1887-1018]{Michael E.~Levi}
\affiliation{Lawrence Berkeley National Laboratory, 1 Cyclotron Road, Berkeley, CA 94720, USA}

\author{C.~Magneville}
\affiliation{IRFU, CEA, Universit\'{e} Paris-Saclay, F-91191 Gif-sur-Yvette, France}

\author{Marc Manera}
\affiliation{Institut de F\'{i}sica d’Altes Energies (IFAE), The Barcelona Institute of Science and Technology, Campus UAB, 08193 Bellaterra Barcelona, Spain}
\affiliation{Serra H\'{u}nter Fellow, Departament de F\'{i}sica, Universitat Au\`{o}noma de Barcelona, Bellaterra, Spain}

\author[0000-0002-4279-4182]{P.~Martini}
\affiliation{Center for Cosmology and AstroParticle Physics, The Ohio State University, 191 West Woodruff Avenue, Columbus, OH 43210, USA}
\affiliation{Department of Astronomy, The Ohio State University, 4055 McPherson Laboratory, 140 W 18th Avenue, Columbus, OH 43210, USA}

\author[0000-0002-1125-7384]{Aaron M. Meisner}
\affiliation{NSF's National Optical-Infrared Astronomy Research Laboratory, 950 N. Cherry Avenue, Tucson, AZ 85719, USA}

\author{R.~Miquel}
\affiliation{Instituci\'{o} Catalana de Recerca i Estudis Avan\c{c}ats, Passeig de Llu\'{\i}s Companys, 23, 08010 Barcelona, Spain}
\affiliation{Institut de F\'{i}sica d’Altes Energies (IFAE), The Barcelona Institute of Science and Technology, Campus UAB, 08193 Bellaterra Barcelona, Spain}

\author[0000-0002-2733-4559]{J.~Moustakas}
\affiliation{Department of Physics and Astronomy, Siena College, 515 Loudon Road, Loudonville, NY 12211, USA}

\author{Adam~D.~Myers}
\affiliation{Department of Physics \& Astronomy, University  of Wyoming, 1000 E. University, Dept.~3905, Laramie, WY 82071, USA}

\author{Jundan~Nie}
\affiliation{National Astronomical Observatories, Chinese Academy of Sciences, A20 Datun Rd., Chaoyang District, Beijing, 100101, P.R. China}

\author{N.~Palanque-Delabrouille}
\affiliation{Lawrence Berkeley National Laboratory, 1 Cyclotron Road, Berkeley, CA 94720, USA}
\affiliation{IRFU, CEA, Universit\'{e} Paris-Saclay, F-91191 Gif-sur-Yvette, France}

\author{W.~J.~Percival}
\affiliation{Waterloo Centre for Astrophysics, University of Waterloo, 200 University Ave W, Waterloo, ON N2L 3G1, Canada}
\affiliation{Department of Physics and Astronomy, University of Waterloo, 200 University Ave W, Waterloo, ON N2L 3G1, Canada}
\affiliation{Perimeter Institute for Theoretical Physics, 31 Caroline St. North, Waterloo, ON N2L 2Y5, Canada}

\author{C.~Poppett}
\affiliation{Lawrence Berkeley National Laboratory, 1 Cyclotron Road, Berkeley, CA 94720, USA}
\affiliation{Space Sciences Laboratory, University of California, Berkeley, 7 Gauss Way, Berkeley, CA  94720, USA}
\affiliation{University of California, Berkeley, 110 Sproul Hall \#5800 Berkeley, CA 94720, USA}

\author{F.~Prada}
\affiliation{Instituto de Astrofisica de Andaluc\'{i}a, Glorieta de la Astronom\'{i}a, s/n, E-18008 Granada, Spain}

\author[0000-0001-5999-7923]{A.~Raichoor}
\affiliation{Lawrence Berkeley National Laboratory, 1 Cyclotron Road, Berkeley, CA 94720, USA}

\author{A.~J.~Ross}
\affiliation{Center for Cosmology and AstroParticle Physics, The Ohio State University, 191 West Woodruff Avenue, Columbus, OH 43210, USA}

\author[0000-0002-3569-7421]{E.~Schlafly}
\affiliation{3700 San Martin Drive, Baltimore, MD 21218, USA}

\author{D.~Schlegel}
\affiliation{Lawrence Berkeley National Laboratory, 1 Cyclotron Road, Berkeley, CA 94720, USA}

\author{M.~Schubnell}
\affiliation{Department of Physics, University of Michigan, Ann Arbor, MI 48109, USA}
\affiliation{University of Michigan, Ann Arbor, MI 48109, USA}

\author{Gregory~Tarl\'{e}}
\affiliation{University of Michigan, Ann Arbor, MI 48109, USA}

\author{B.~A.~Weaver}
\affiliation{NSF's National Optical-Infrared Astronomy Research Laboratory, 950 N. Cherry Avenue, Tucson, AZ 85719, USA}

\author{R.~H.~Wechsler}
\affiliation{Kavli Institute for Particle Astrophysics and Cosmology, Stanford University, Menlo Park, CA 94305, USA}
\affiliation{Physics Department, Stanford University, Stanford, CA 93405, USA}
\affiliation{SLAC National Accelerator Laboratory, Menlo Park, CA 94305, USA}

\author{Christophe~Yèche}
\affiliation{IRFU, CEA, Universit\'{e} Paris-Saclay, F-91191 Gif-sur-Yvette, France}

\author{Zhimin~Zhou}
\affiliation{National Astronomical Observatories, Chinese Academy of Sciences, A20 Datun Rd., Chaoyang District, Beijing, 100101, P.R. China}

\begin{abstract}
The Dark Energy Spectroscopic Instrument (DESI) is carrying out a 5-year survey that aims to measure the redshifts of tens of millions of galaxies and quasars, including 8 million luminous red galaxies (LRGs) in the redshift range of $0.4<z<{\sim}\,1.0$. Here we present the selection of the DESI LRG sample and assess its spectroscopic performance using data from Survey Validation (SV) and the first 2 months of the Main Survey. The DESI LRG sample, selected using $g$, $r$, $z$, and $W1$ photometry from the DESI Legacy Imaging Surveys, is highly robust against imaging systematics. The sample has a target density of 605 deg$^{-2}$ and a comoving number density of $5\times10^{-4}\ h^3\mathrm{Mpc}^{-3}$ in $0.4<z<0.8$; this is a significantly higher density than previous LRG surveys (such as SDSS, BOSS and eBOSS) while also extending to $z \sim 1$. After applying a bright star veto mask developed for the sample, $98.9\%$ of the observed LRG targets yield confident redshifts (with a catastrophic failure rate of $0.2\%$ in the confident redshifts), and only $0.5\%$ of the LRG targets are stellar contamination. The LRG redshift efficiency varies with source brightness and effective exposure time, and we present a simple model that accurately characterizes this dependence. In the appendices, we describe the extended LRG samples observed during SV.
\end{abstract}


\section{Introduction}

Galaxy redshift surveys have been established as a pillar of observational cosmology over the past several decades. The large-scale structures, traced by galaxies, reveal the imprint of baryon acoustic oscillations (BAO), a feature that can be used to measure the expansion history of the Universe. The redshift space distortions (RSD) caused by the peculiar velocities of galaxies enable measurements of the growth of the large-scale structure and tests of general relativity.

Luminous red galaxies (LRGs) are an important type of galaxies for large-area redshift surveys, and are specifically selected for observations due to two main advantages: 1) they are bright galaxies with the prominent $4000\,\text{\AA}$ break in their spectra, thus allowing for relatively easy target selection and redshift measurements; and 2) they are highly biased tracers of the large-scale structure, thus yielding a higher S/N per-object for the BAO measurement compared to typical galaxies. In addition to the cosmological constraints from BAO and RSD, there will be significant gains in constraining powers when the LRG sample is combined with other observations, e.g., using the LRGs (and their massive dark matter halos) as gravitational lenses of background galaxies and the cosmic microwave background (e.g., \citealt{mandelbaum_cosmological_2013,singh_cosmological_2020,white_cosmological_2022}).

\begin{figure*}
    \centering
    \includegraphics[width=1.4\columnwidth]{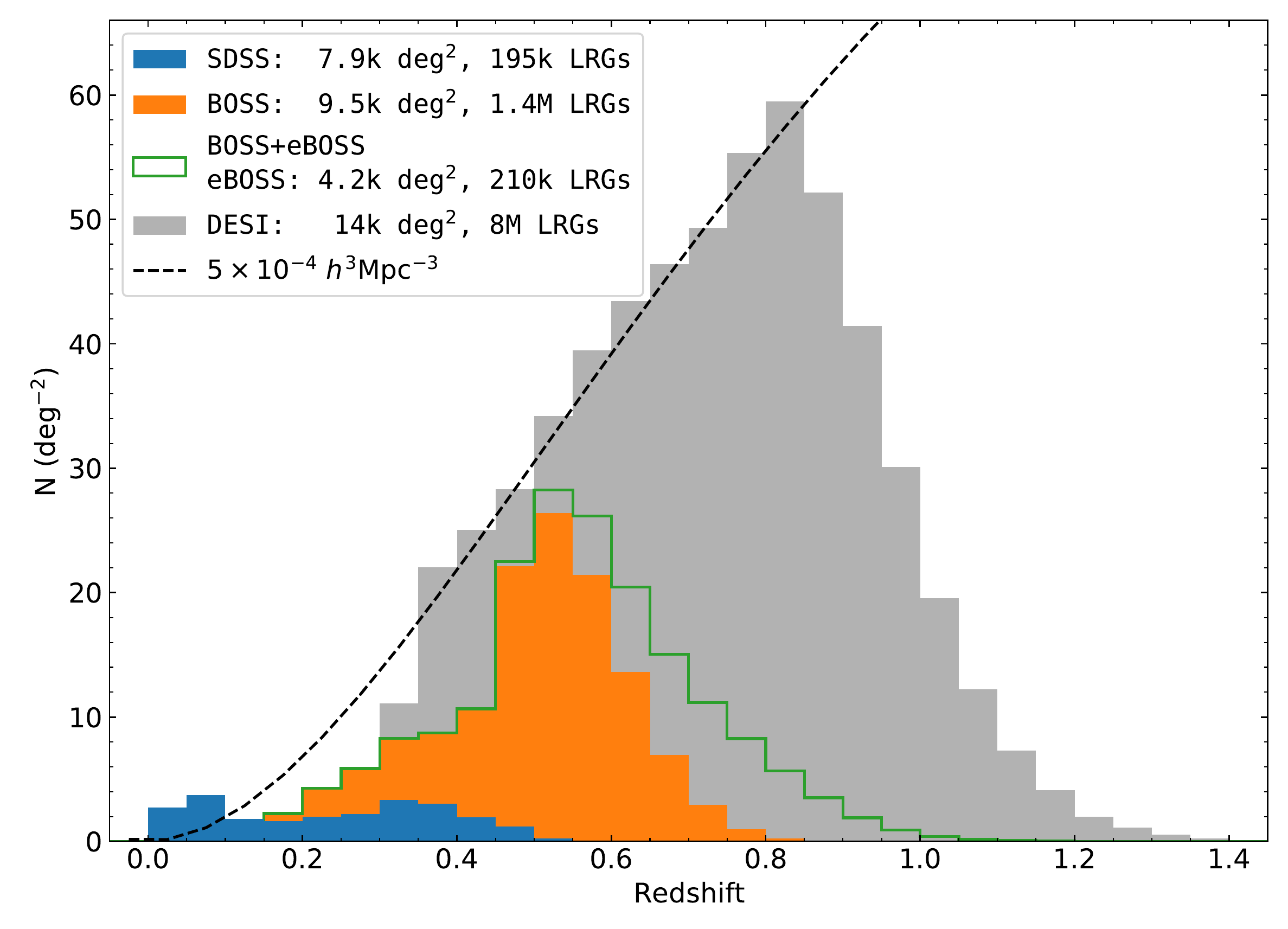}
    \caption{The redshift distribution of the DESI LRG sample and its comparison with LRG samples from earlier surveys. The y-axis is the number of objects in each redshift bin (of width $\Delta z=0.05$) per deg$^2$. The survey area and the total number of LRGs that have been or will be observed in each survey are listed in the legend. The dashed curve corresponds to the redshift distribution of a hypothetical sample with a constant comoving density of $5\times10^{-4}\ h^3\mathrm{Mpc}^{-3}$, which is approximately the DESI LRG target density in the redshift range of $0.4<z<0.8$; the area under the curve is proportional to the enclosed comoving volume. We describe how we obtain the redshifts for DESI LRGs in \S\ref{sec:spectro_data}.}
    \label{fig:main_dndz}
\end{figure*}

The Dark Energy Spectroscopic Instrument (DESI; \citealt{desi_collaboration_desi_2016, desicollaboration_desi_2016, desi_instrument_overview}) is undertaking the largest galaxy redshift survey to date, and LRGs will be the primary galaxy targets that DESI will observe in the redshift range of $0.4<z<{\sim}\,1.0$. Compared to LRG samples from previous surveys, such as the SDSS LRG survey \citep{eisenstein_spectroscopic_2001,eisenstein_detection_2005}, BOSS \citep{reid_sdssiii_2016,alam_clustering_2017} and eBOSS \citep{prakash_sdssiv_2016,ebosscollaboration_completed_2021}, the DESI LRG sample has a significantly higher target density and extends to higher redshifts (see Figure \ref{fig:main_dndz}). This is made possible by DESI's higher fiber multiplexing, larger telescope aperture and better spectroscopic performance, and the availability of deeper (and highly uniform) imaging data necessary for target selection.

In this paper, we describe the selection of the DESI LRG targets and assess the selection uniformity and spectroscopic performance. Significant efforts were made to minimize the impact of imaging systematics (i.e. the modulation of the target density caused by image quality variations across the sky). These include improvements to the image reduction pipeline, which were motivated by the need for uniform target selection and are discussed in \citet{schlegel_dr9}, as well as making careful choices for target selection. We will describe these choices for the DESI LRG sample in this paper.

The structure of the paper is as follows. We describe the imaging data, selection cuts, stellar mass completeness, and veto masks in \S\ref{sec:target_selection}. We assess potential imaging systematics in \S\ref{sec:ts_systematics}. In \S\ref{sec:spectro_assessment}, we evaluate the spectroscopic redshift efficiency and model its dependence on source brightness and exposure time. We summarize our work in \S\ref{sec:summary}. In the appendices, we describe the selections and redshift performance of the extended LRG samples observed before the start of the Main Survey, specifically during Survey Validation (``SV1'') and the 1\% Survey (``SV3''); these observations were done separately and are not included in the Main LRG sample (which we simply refer to as DESI LRGs).

This paper is part of a series of papers presenting the selection of DESI targets and their characterization. \citet{dawson_sv_overview} gives an overview of the DESI targets and the Survey Validation program, and \citet{myers_target_pipeline} describes how the target selection algorithms are implemented in DESI. The Milky Way Survey (MWS) sample is presented in \citet{cooper_mws}, the Bright Galaxy Survey (BGS) sample in \citet{hahn_desi_BGS_selection}, the emission line galaxy (ELG) sample in \citet{raichoor_desi_elg_selection}, and the quasar (QSO) sample in \citet{chaussidon_desi_qso_selection}. \citet{lan_desi_galaxy_vi} and \citet{alexander_desi_qso_vi} describe the creation of spectroscopic truth tables based on visual inspections of the galaxy (BGS, LRG, and ELG) spectra and the QSO spectra, respectively.

\section{Target selection}
\label{sec:target_selection}

\subsection{Imaging data}

\begin{figure*}
    \centering
    \includegraphics[width=2.05\columnwidth]{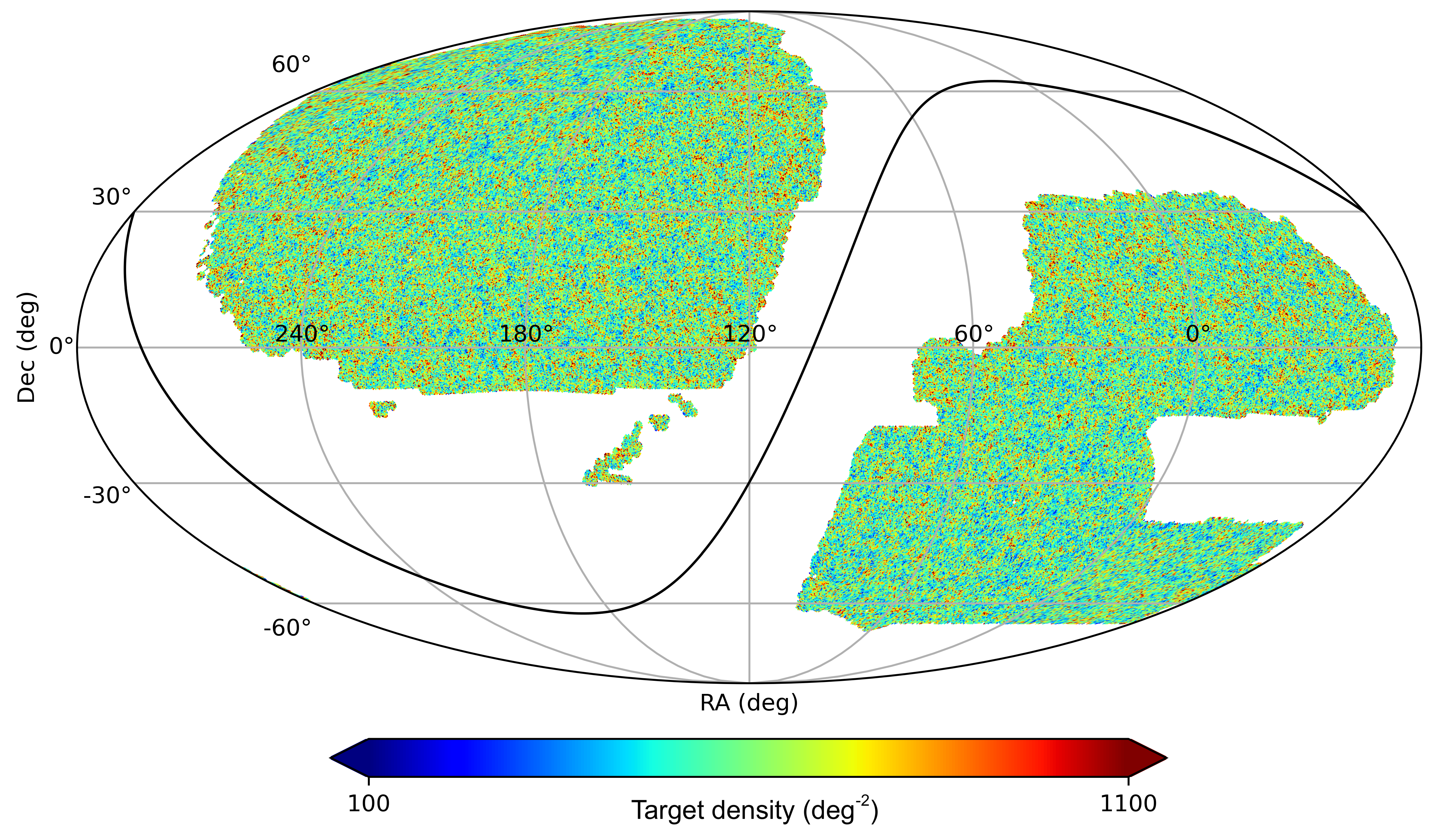}
    \caption{The footprint of the DESI Legacy Imaging Surveys DR9, with the colors representing the surface density (in deg$^{-2}$) of the LRG targets (after applying the LRG veto masks). DESI will only observe regions above DEC$>$-20, and the DESI footprint also avoids regions close to the edge of the imaging footprint. See the actual DESI footprint in \citet{schlafly_survey_ops}. This density map is computed with a HEALPix resolution of NSIDE=256, and we only plot pixels that are $>20\%$ occupied by the imaging survey footprint. The curve that separates the two regions is the Galactic plane.}
    \label{fig:density_map}
\end{figure*}

The LRG targets are selected from the DESI Legacy Imaging Surveys Data Release 9 (\citealt{schlegel_dr9,dey_overview_2019}, hereafter LS DR9), specifically the $g$, $r$, $z$ optical bands ($4000$--$10000\text{\AA}$, without the $i$ band at $7800\text{\AA}$) and forced photometry of the {\it WISE} \citep{wright_widefield_2010} $W1$ band in the infrared ($3.4\,\micron$). The imaging footprint is shown in Figure \ref{fig:density_map}. Table \ref{tab:lrg_summary} lists the sky areas and other summary information about the DESI LRG sample.

\begin{table}
    \centering
    \caption{Useful information about DESI and the LRG targets.}
    \label{tab:lrg_summary}
    \begin{tabular}{ll}
    \hline
    \hline
    Area in the imaging footprint & 19700 deg$^2$ \\
    Area in the DESI survey & 14800 deg$^2$ \\
    Fraction of area in target mask & 1.0\% \\
    Fraction of area in LRG veto mask & 8.5\% \\
    Target density & 605 deg$^{-2}$ \\
    Spectroscopically confirmed star fraction & 0.5\% \\
    Spectroscopically confirmed quasar fraction & 1.6\% \\
    Fraction rejected by redshift quality cut & 1.1\% \\
    Fraction of catastrophic redshift failures & 0.2\% \\
    \hline
    \end{tabular}
    \tablecomments{The areas include all regions with optical ($grz$) coverage without any masking. The area in the DESI survey is approximate. The LRG veto mask includes the target mask. The target density is the average density over the DESI footprint. The target density and all the spectroscopy/redshift-related values are calculated after the LRG veto mask is applied. The catastrophic redshift failure rate is calculated after applying the redshift quality cut.}
\end{table}


The optical $grz$ imaging consists of two regions separated at declination $\delta={\sim}\,32\degr$, with each region observed by different telescopes (with similar $grz$ filter sets). The southern ($\delta<{\sim}\,32\degr$) part of the imaging footprint (hereafter ``the South'') is observed by DECam on the 4m Blanco Telescope at the Cerro Tololo Inter-American Observatory (CTIO). Most of the observation is done by the DECam Legacy Survey (DECaLS, \citealt{dey_overview_2019}), and data from other observations, most importantly the Dark Energy Survey (DES, \citealt{thedarkenergysurveycollaboration_dark_2005}), is also used. The northern ($\delta>{\sim}\,32\degr$) part of the imaging footprint (hereafter ``the North''), which is inaccessible from CTIO, is observed by two telescopes at the Kitt Peak National Observatory in two surveys: the Beijing–Arizona Sky Survey (BASS, \citealt{zou_project_2017}) observed in the $g$ and $r$ bands using the 90Prime Camera on the 2.3m Bok Telescope, and the Mayall $z$-band Legacy Survey (MzLS) observed in the $z$ band using the Mosaic-3 Camera on the 4m Mayall Telescope. The Mayall Telescope has since been repurposed for DESI.

For the same photometric band, small differences in the filter sets, detectors, observing conditions, and image processing between the North and the South cause their photometry to differ slightly. For the LRGs, these differences are often non-linear functions of magnitude and color, and cannot be quantified by a constant offset. Thus we find it necessary to implement slightly different color cuts for each region. The cuts are optimized using both photometric and spectroscopic data to achieve uniform density and redshift distributions across the footprint. The specific color cuts are described in the next subsection.

\subsection{Selection cuts}
\label{sec:selection_cuts}

\begin{figure*}
    \centering
    \includegraphics[width=1.55\columnwidth]{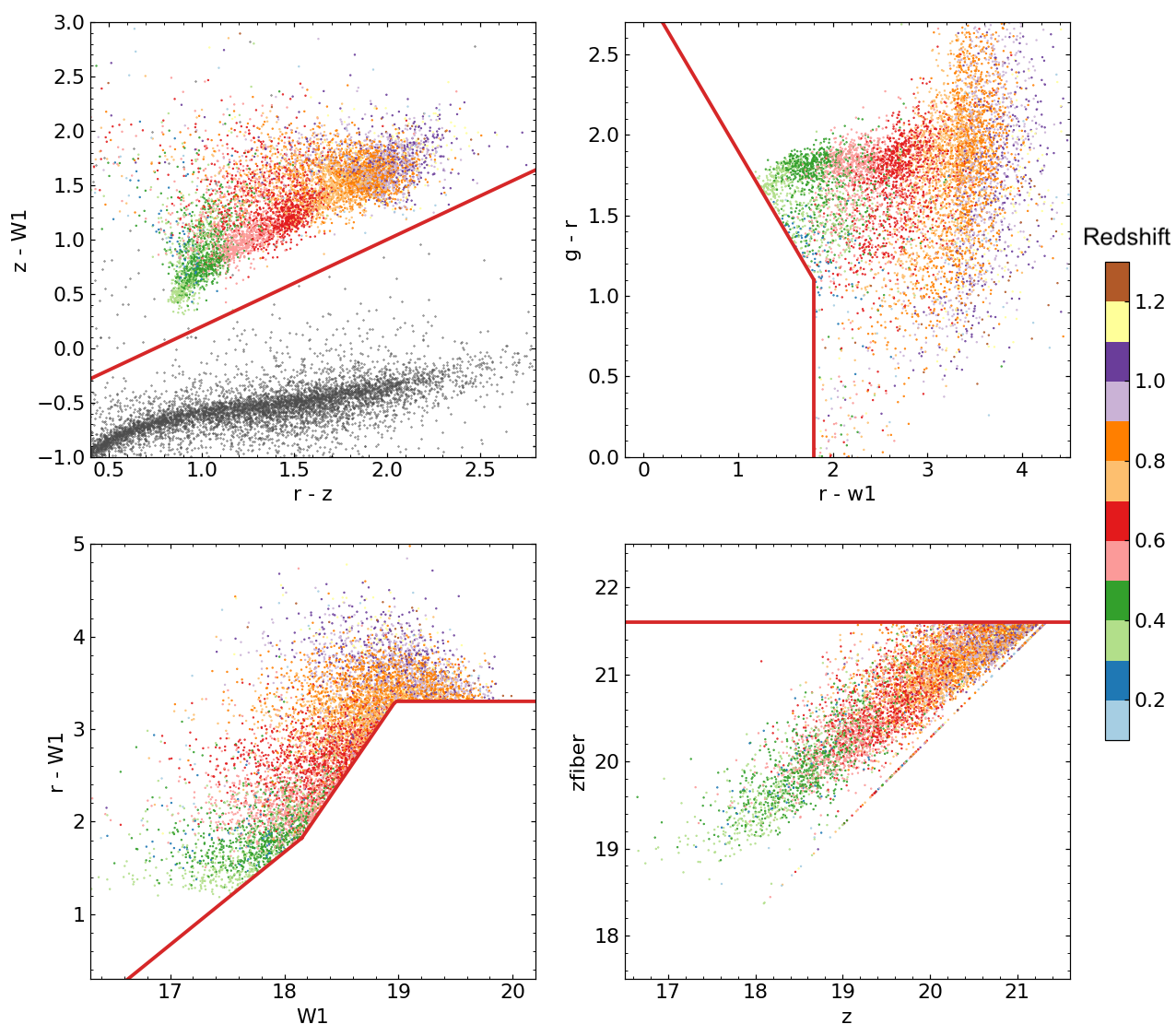}
    \caption{Selection cuts for the LRG targets in the South footprint.
    The points are color-coded by their redshifts measured by DESI. The upper left panel shows the stellar rejection cut, with gray points representing stars (which are plotted to show the stellar locus and are not LRG targets). The upper right panel shows the cut that removes lower-redshift and bluer galaxies. The lower left panel shows the sliding color-magnitude cut that serves as the luminosity cut and also shapes the redshift distribution; the ``knee'' at $W1={\sim}\,19$ introduces more galaxies at higher redshift. The lower right panel shows the magnitude limit in $z$-band fiber magnitude that ensures enough S/N for DESI observations.}
    \label{fig:main_lrg_selection}
\end{figure*}

The LRG targets are selected using optical photometry in the $grz$ bands and near-infrared photometry in the {\it WISE} $W1$. The LRG selection cuts for the South, shown in Figure \ref{fig:main_lrg_selection}, are

\begin{subequations}
\label{eq:lrg_selection_south}
\begin{align}
    & z_\mathrm{fiber} < 21.60 \label{eq:mag-limit}\\
    & z-W1 > 0.8\times(r-z) - 0.6 \label{eq:non-stellar}\\
    & (g-W1>2.9) \ \mathrm{OR}\  (r-W1>1.8)  \label{eq:low-z}\\
   \begin{split}
    & ((r-W1 > 1.8\times(W1-17.14)) \ \mathrm{AND} \\
    & (r-W1 > W1-16.33)) \ \mathrm{OR}\ (r-W1>3.3) \label{eq:sliding-cut}
    \end{split}
\end{align}
\end{subequations}

where $g$, $r$, $z$, and $W1$ are magnitudes and $z_\mathrm{fiber}$ is the $z$-band fiber magnitude, i.e., the magnitude corresponding to the expected flux within a DESI fiber\footnote{It uses the FIBERFLUX\_Z value in the imaging data; see \url{https://www.legacysurvey.org/dr9/files/\#sweep-catalogs-region-sweep}}. Throughout this paper, unless otherwise specified, all the magnitudes are in the AB system and are corrected for Galactic extinction\footnote{See the derivation of the extinction coefficients described here: \url{https://www.legacysurvey.org/dr9/catalogs/\#galactic-extinction-coefficients}} based on \citealt{schlegel_maps_1998} with correction from \citealt{schlafly_measuring_2011}.

Eqn. \ref{eq:non-stellar} utilizes the $1.6$\,\micron\ (restframe)``bump'' \citep{John88,Sawicki02} to efficiently remove stars from the sample (as shown in the upper left panel of Figure \ref{fig:main_lrg_selection}), similar to the stellar-rejection cut in \citet{prakash_luminous_2015}, resulting in a low stellar contamination rate of ${\sim}\,0.5\%$ (see \S\ref{sec:spectro_classification}). Eqn. \ref{eq:low-z} removes galaxies at lower redshifts while retaining high completeness of massive galaxies at $z>0.4$.
The sliding color-magnitude cuts in Eqn. \ref{eq:sliding-cut} function as luminosity cuts: as shown in the lower left panel of Figure \ref{fig:main_lrg_selection}, the $r-W1$ color is a good proxy for redshift, and the $W1$ magnitude limit that shifts with $r-W1$ effectively selects the most luminous (in the observed $W1$ band) galaxies at any redshift. For objects with $r-W1>{\sim}\,3.3$, which are the faintest LRGs in our sample and are near the faint limit (Eqn. \ref{eq:mag-limit}), the $r-W1$ vs $W1$ sliding cut is dropped to boost the number density of the highest-redshift LRGs. The cuts are tuned so that the comoving number density of the selected sample is close to constant at $0.4<z<0.8$. 

Finally, eqn. \ref{eq:mag-limit} sets the faint limit for the sample to ensure a high redshift success rate for DESI spectroscopic observations (see \S\ref{sec:z_quality_cut}-\ref{sec:depth_and_mag_dependence}). We choose the fiber magnitude over the total magnitude as the faint limit because the former is much more strongly correlated with the spectroscopic S/N. Note that the fiber magnitude cut, which is effectively a morphology cut, could introduce systematics in both sample selection (e.g., systematics dependence on seeing) and theoretical modeling (as galaxy orientation could be aligned with the tidal field; e.g., see \citealt{lamman_intrinsic_2022}), and such effects will need to be carefully studied in the clustering analysis.

The photometry in the North is slightly different from the South, and the selection cuts are tuned to match the number density and the redshift distribution in the South. The cuts for the North are

\begin{subequations}
\label{eq:lrg_selection_north}
\begin{align}
    & z_\mathrm{fiber} < 21.61 \label{eq:mag-limit-north}\\
    & z-W1 > 0.8\times(r-z) - 0.6 \\
    & (g-W1>2.97) \ \mathrm{OR}\  (r-W1>1.8) \label{eq:low-z-north}\\
   \begin{split}
    & ((r-W1 > 1.83\times(W1-17.13)) \ \mathrm{AND} \\
    & (r-W1 > W1-16.31))\ \mathrm{OR}\ (r-W1>3.4)
    \end{split}
\end{align}
\end{subequations}

In addition to the aforementioned cuts, we apply the following quality cuts (e.g. to remove objects with bad photometry) in the target selection pipeline \citep{myers_target_pipeline}. We require that each object be observed at least once in all of the three optical bands. We also require that the inverse-variance values for $r$, $z$ and $W1$ fluxes be positive; this rejects problematic imaging data. A small number of stars are not removed by the aforementioned selection cuts due to saturation in the imaging. We remove them by requiring $z_\mathrm{fibertot}>17.5$, and if Gaia \citep{gaiacollaboration_gaia_2018} photometry is available, we also require $G_\mathrm{Gaia}>18$.

The DESI target selection pipeline also removes objects close to bright stars (in Gaia and Tycho-2), star clusters, and large galaxies, which are flagged by the LS DR9 MASKBITS\footnote{\url{https://www.legacysurvey.org/dr9/bitmasks/\#maskbits}} 1, 12, and 13, respectively. These masks are very minimal and only remove regions with the worst contamination, and additional masking is needed for the clustering analysis. We describe these additional masks in \S\ref{sec:masks}.

The selection cuts for the Survey Validation and the 1\% Survey samples are described in Appendices \S\ref{sec:sv1} and \S\ref{sec:sv3}, respectively.

\subsection{Stellar Mass Completeness}

In order to accurately model galaxy clustering (e.g. \citealt{Zu2015clustering,zhou_clustering_2021}) and study galaxy-galaxy lensing (e.g. \citealt{Alam2017lensing,Jullo2019lensing}), halo occupation distribution (e.g. \citealt{RodriguezTorres2016HOD}) and evolution of the most massive galaxies (e.g. \citealt{Bundy2017MassiveGalaxies}), it is desirable to have a large spectroscopic sample of strongly clustered galaxies with well defined stellar populations. The target selection cuts for the DESI LRG sample have therefore been optimized to select the most massive galaxies with a high degree of completeness. We define completeness as
the ratio of selected LRGs to the total number of galaxies brighter than the LRG magnitude limit
(defined by eqns.~\ref{eq:mag-limit}~\&~\ref{eq:mag-limit-north}). In this section we refer to objects that satisfy our stellar rejection cut (defined by eq.~\ref{eq:non-stellar}) as ``galaxies''.

The cuts defined by Eqns. ~\ref{eq:low-z} \& \ref{eq:low-z-north} reject objects with low redshifts while retaining the most massive galaxies for redshifts over 0.4. The design of these cuts was guided by estimates of stellar masses of galaxies obtained using a random forest algorithm \citep{Breiman2001RandomForest} trained on DESI Legacy Imaging Survey photometry and stellar masses of galaxies from \citet{Bundy2015Stripe82Catalog}. A detailed description of the method used to obtain the stellar masses can be found in appendix~\ref{app:rf_masses}.

Figure \ref{fig:mass_completeness} shows the stellar mass completeness of the DESI LRG sample both as a function of stellar mass and redshift. As spectroscopic redshifts are only available for some of the selected LRGs and not the magnitude-limited sample, we use the photometric redshifts in LS DR9\footnote{\url{https://www.legacysurvey.org/dr9/files/\#photometric-redshift-sweeps}} \citep{zhou_clustering_2021} for this analysis. The selection cuts result in a sample which is highly complete for the most massive galaxies (i.e. $\log_{10}(\mathrm{M}_{\star}[\mathrm{M}_{\odot}])>11.5$) in the redshift range of 0.4 to 1.0. The completeness decreases significantly for redshifts lower than 0.4 but the decrease is less severe for redshifts above 1.0. This high mass completeness is one of the defining characteristics of the DESI LRG sample and will aid a multitude of scientific studies.

\begin{figure*}
    \centering
    \includegraphics[width=1.95\columnwidth]{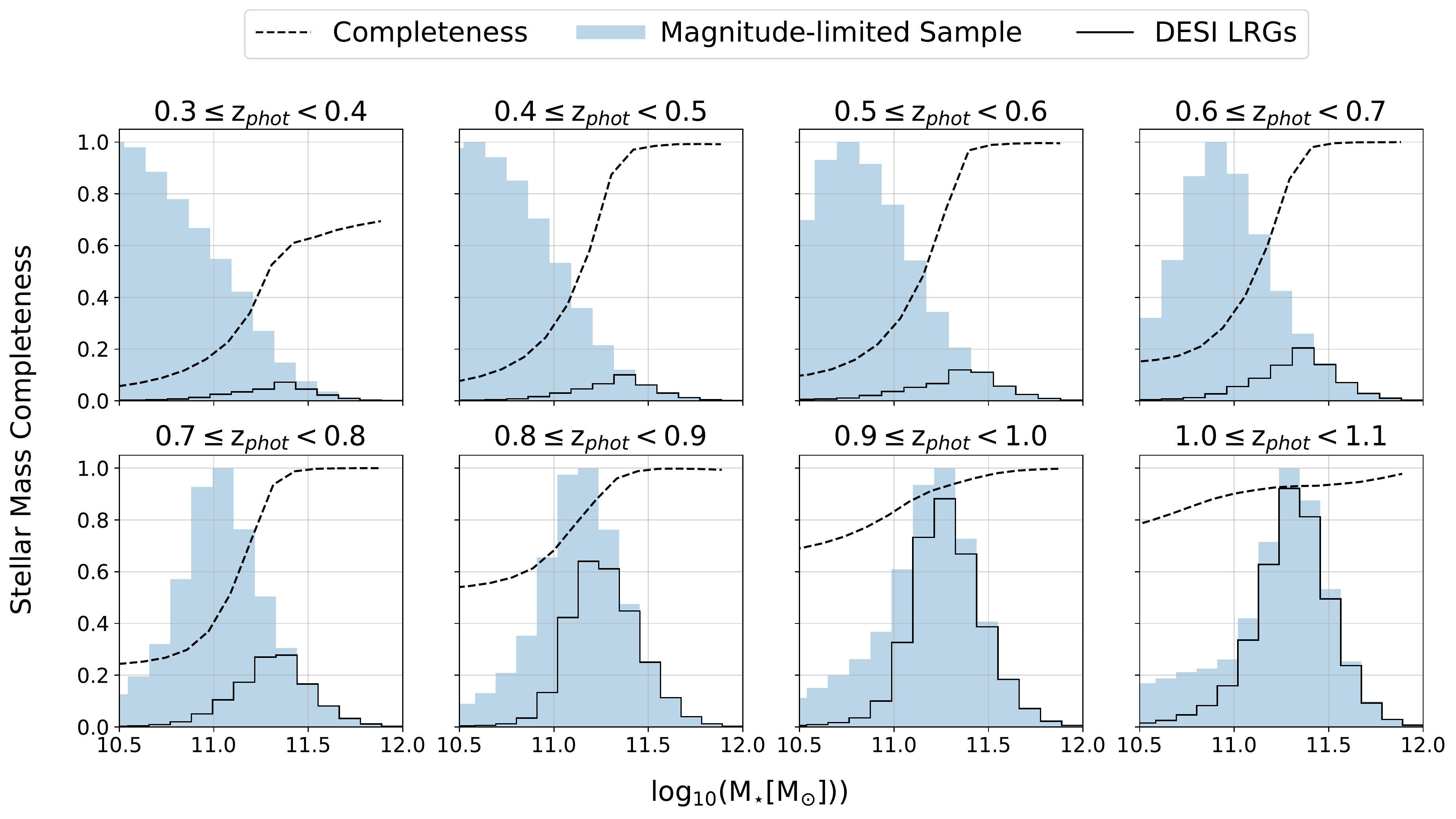}
    \caption{Stellar mass completeness of the DESI LRG sample as a function of stellar mass and photometric redshift. The dashed curve shows the fraction of galaxies above a given stellar mass that has been selected as a DESI LRG compared to a magnitude-limited sample. The blue histogram shows the distribution of stellar masses of a magnitude-limited sample of galaxies (having the same magnitude limit as the DESI LRG sample) whereas the black histogram denotes the subset of galaxies that have been selected as DESI LRGs. The stellar masses were obtained using a random forest-based algorithm (described in appendix~\ref{app:rf_masses}) and the photometric redshifts are from \citet{zhou_clustering_2021}. As spectroscopic redshifts are not available for the magnitude-limited sample, we use photometric redshifts for this demonstration. The figure uses objects from both the North and the South where valid photometry in $g$, $r$, $z$, $\mathrm{W}_{1}$ and $\mathrm{W}_{2}$ is available. The selected sample is highly complete for the most massive galaxies (i.e. $\log_{10}(\mathrm{M}_{\star}[\mathrm{M}_{\odot}])>11.5$) in the redshift range of 0.4 to 1.0. The completeness decreases significantly for redshifts lower than 0.4 but the decrease is less steep for redshifts above 1.0.}
\label{fig:mass_completeness}
\end{figure*}

\subsection{Veto masks for clustering analysis}
\label{sec:masks}

Here we describe the additional veto masks specifically optimized for the LRG targets to create a clean sample for the DESI clustering analysis. They are mostly masks of bright stars, although they also include masking for large galaxies and star clusters, etc. The veto masks are comprised of four separate sets of masks:

1. unWISE \citep{meisner_unwise_2019} pixel-level bitmask: we use all but bit 5 (``AllWISE-like circular halo'') of the collapsed mask bits as listed in Table A4 of \citet{meisner_unwise_2019}. We exclude bit 5 because these circular masks are not optimal (either too large or too small, depending on the magnitude of the star) for the LRG targets, and they are replaced by

2. WISE circular geometric masks: these masks replace the ``AllWISE-like circular halo'' masks in unWISE. The radius vs W1 magnitude relation is optimized for the LRGs, so that the excess or deficit of LRG targets at the edge of the mask is less than ${\sim}\,10\%$.

3. Gaia/Tycho-2 circular masks: similarly, we obtain the radius vs magnitude relation for Gaia stars. We use stars in Gaia Early Data Release 3 (EDR3, \citealt{gaia_edr3}) supplemented at the bright end (where Gaia photometry is unreliable) by Tycho-2 and 2MASS photometry.

4. Custom masks: these are masks for large galaxies, star clusters, and planetary nebulae that were not masked by the LS DR9 MASKBITS, and regions with other imaging artifacts (identified from visual inspection of regions with high LRG densities). The total area of the custom masks is much smaller than the other masks.

We describe 2-4 in more detail in Appendix \ref{sec:appdx_masks}. The combined veto masks remove 8.5\% of the DESI footprint. Within the masked (contaminated) area, the LRG target density is 1100 deg$^{-2}$ and the stellar contamination rate (based on spectroscopic classification) is ${\sim}\,10\%$, compared to the 605 deg$^{-2}$ density and ${\sim}\,0.5\%$ stellar contamination in the unmasked (``clean'') area. The stellar contamination rate is much higher in the masked region because the photometry (especially in the $W1$ band) used in the stellar rejection cut (eqn. \ref{eq:non-stellar}) is contaminated by the nearby bright stars.

While the full LRG spectroscopic and target catalogs include objects flagged by the LRG veto masks, we recommend that those objects be removed for analyses that require a clean sample with uniform selection. In the rest of the paper, we only use the ``clean'' LRG sample (with the LRG veto masks applied) instead of the full target/spectroscopic sample.

Finally, we note that the LRG veto masks presented here are not definitive, and they may see further improvements, e.g., for the DESI Year 1 science analyses.


\section{Target selection systematics}
\label{sec:ts_systematics}


\subsection{Imaging and foreground systematics}

The variation of imaging properties (such as depth and seeing) over the footprint and the presence of astrophysical foregrounds (such as Galactic dust) can imprint on the density of galaxy targets. Here we examine the impact of these systematics on the LRG target density. The systematics properties that we examine here include depth (galaxy depth in $grz$ and PSF depth in $W1$), seeing (in $grz$), Gaia stellar density, and Galactic extinction $E(B-V)$ (from \citealt{schlegel_maps_1998}).

While the photometry used in target selection has been corrected for Galactic extinction, we include $E(B-V)$ here because imperfections in the correction, e.g., due to errors in the dust map, can still bias the photometry and affect the target density.

We use the {\tt STARDENS} values from \citealt{myers_target_pipeline} for Gaia stellar density, and we use values from the imaging catalog\footnote{\url{https://www.legacysurvey.org/dr9/files/\#sweep-catalogs-region-sweep}} for all other systematics properties (GALDEPTH, PSFDEPTH, PSFSIZE and EBV).

\begin{figure*}
    \centering
    \includegraphics[width=2.05\columnwidth]{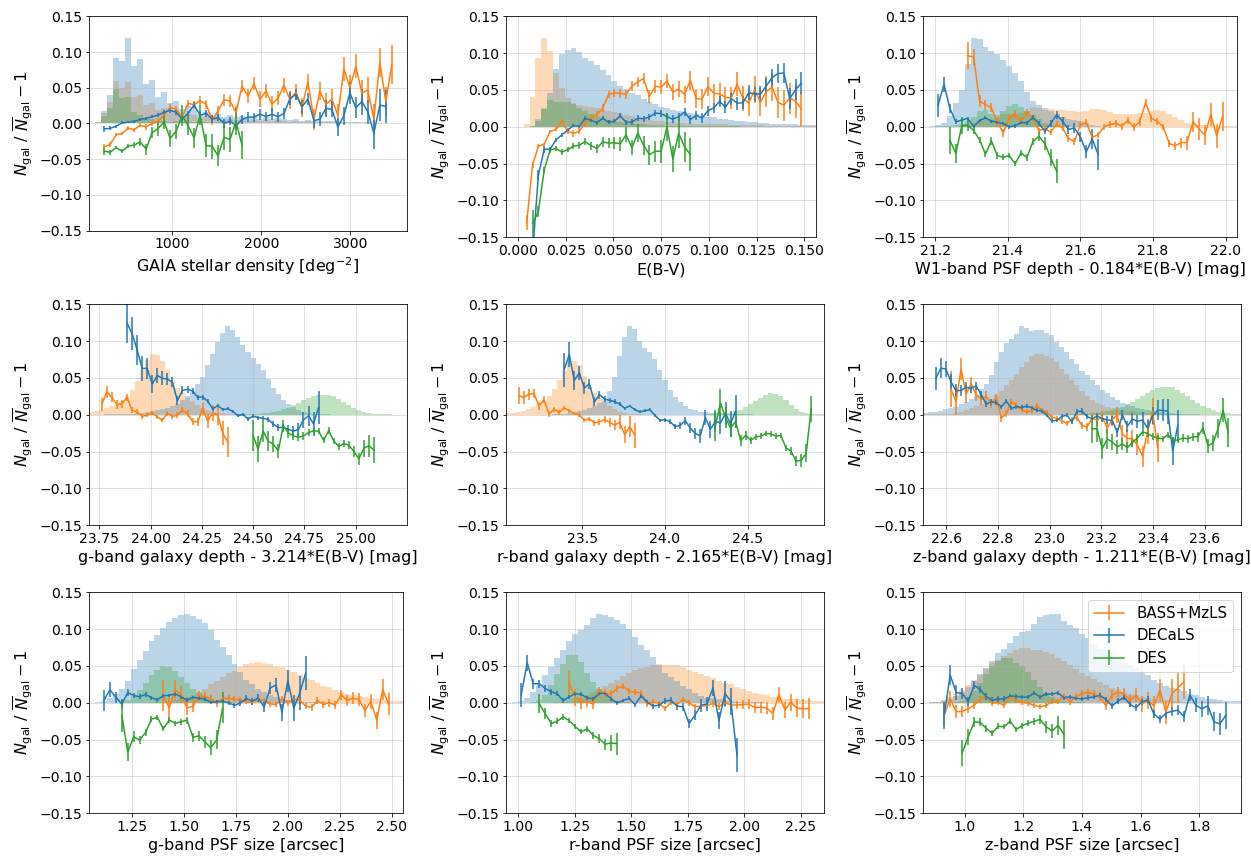}
    \caption{Density of LRG targets in bins of imaging/foreground systematics values in the three imaging regions. (DECaLS and DES are both observed with DECam and have the same selection cuts and linear regression coefficients, but DES is significantly deeper and we plot it as a separate region to illustrate the difference.) The error bars represent ``the error of the mean'' assuming Gaussian distribution. The histograms show the distribution of each systematics property for each imaging region. ``Galaxy depth'' is based on `the `GALDEPTH'' value in the LS DR9 catalog and it is the $5\sigma$ detection magnitude of an ELG-like galaxy and it assumes zero Galactic extinction; to account for Galactic extinction, we add an $E(B-V)$ term to obtain the imaging depth relevant for extragalactic sources. ``PSF size'' is the PSF FWHM and measures the seeing. The trends are computed using a HEALPix density map with NSIDE=512 by averaging over the pixels in bins of imaging/foreground properties.}
    \label{fig:systematics_trends}
\end{figure*}

Figure \ref{fig:systematics_trends} shows the dependence of LRG target density on the imaging and foreground systematics. The LRG sample is much brighter than the imaging detection limit (with the faintest LRG targets being at least ${\sim}\,2$ magnitudes brighter than the median $z$-band 5$\sigma$ detection limit), and the stellar-rejection cut and the LRG veto masks efficiently remove the contamination caused by stars. Therefore the LRG sample is relatively robust against these imaging/foreground systematics. The density deviations caused by these systematics are almost all within $\pm5\%$.

\begin{figure*}
    \centering
    \includegraphics[width=2.05\columnwidth]{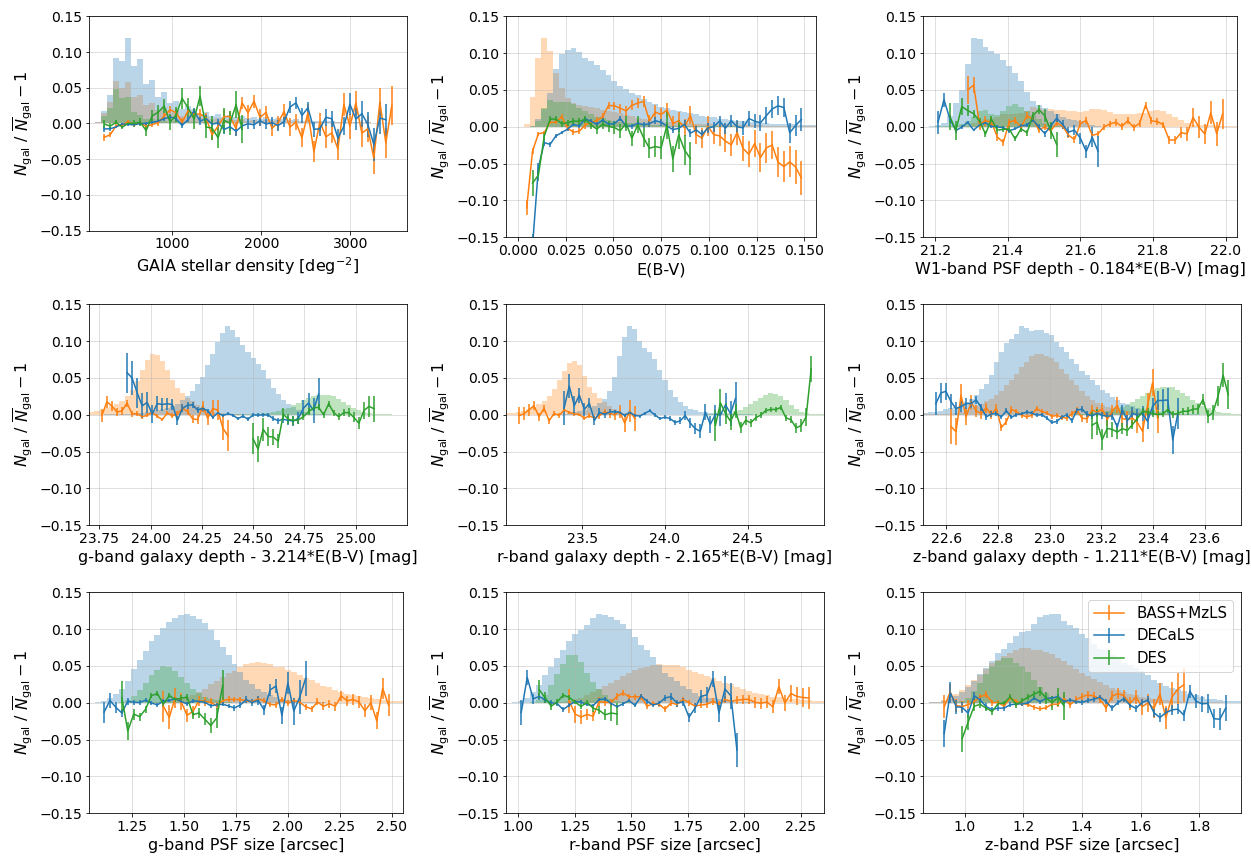}
    \caption{As Figure \ref{fig:systematics_trends} but with linear regression weights applied. The stellar density is not included in the parameters for the linear regression.}
    \label{fig:systematics_trends_corrected}
\end{figure*}

The systematics trends in Figure \ref{fig:systematics_trends} can be almost completely removed via linear regression of the systematics properties:
\begin{equation}
N_\mathrm{predict, k} = c_0 + \sum_{i=1}^{8} \sum_{l=1}^{N_\mathrm{rand, k}} c_i S_{i,l} ,
\end{equation}
where $N_\mathrm{predict, k}$ is the ``predicted'' number of LRG targets in the $k$-th HEALPix pixel, $S_{i, l}$ is the value of the $i$-th systematics property of the $l$-th random, $N_\mathrm{rand, k}$ is the number of randoms in the $k$-th pixel, and $c_i$ are the coefficients. We use the randoms in LS DR9. The ``corrected'' systematics trends are shown in Figure \ref{fig:systematics_trends_corrected}. For the linear regression, we include all systematics properties except stellar density, because 1) the stellar contamination rate is already very low and 2) the stellar density maps are inherently noisy and the Gaia catalog (on which the stellar density map is based) are much brighter than the stellar contamination in the LRG targets. Indeed we find little correlation with stellar densities after applying the systematics weights. The coefficients for the North and the South are different, but DECaLS and DES are treated as one sample in the linear regression and have the same coefficients. The LRG density in the DES region is ${\sim}\,3\%$ lower than the average density due to its deeper photometry, and the linear regression weights accurately predict this density difference.

There is an unexpected dependence on $E(B-V)$ that remains after correcting for depth and seeing dependence: the LRG density at $E(B-V) \lesssim 0.015$ is more than $5\%$ lower than average. We also see similar trends at very low $E(B-V)$ in the other DESI tracers (which have very different selections and redshifts), so it is unlikely to be a statistical fluke. While we are not certain what causes this drop in target density at very low $E(B-V)$, we speculate that it is caused by systematics in the Galactic extinction map. Specifically, the $E(B-V)$ map from \citet{schlegel_maps_1998} is based on dust emission in the far infrared, which may have been contaminated by FIR emissions from background galaxies. The remaining trends (at higher $E(B-V)$) might be due to SED-dependent effects: ideally, we would calculate the extinction coefficients for each galaxy based on its spectral energy distribution (SED), but in practice, we use a single stellar spectrum for calculating the extinction coefficients, and this could lead to systematic errors in the dereddened fluxes. We will examine the Galactic dust-related systematics in future investigations.

Note that while the aforementioned linear weights work well for the projected density of the full LRG sample, for subsets of the sample (e.g., in redshift bins) the coefficients and weights should be recomputed for each subset, because different subsamples are affected by the selection cuts differently (e.g., brighter subsamples might be more sensitive to the sliding cut in $r-W1$ vs $W1$ while fainter subsamples might be more sensitive to the $z_\mathrm{fiber}$ faint limit) and have different sensitivities to the different systematics.

There were no formal requirements on DESI targeting w.r.t. these systematic trends in target density. Instead, this analysis was used as a diagnostic to ensure no trends existed beyond expectations, e.g., from previous surveys. One can observe that the trends are generally more moderate than those found in BOSS and eBOSS \citep{ross_clustering_2017,ross_completed_2020}. Further, we have demonstrated that a simple linear regression, similar again to that applied to BOSS/eBOSS, successfully models the trends. The exception is a relationship with $E(B-V)$ that should be studied in more detail in future DESI work. The details and modeling here thus represent a first step in the process of determining the selection function for DESI LRG LSS catalogs, a vital component for producing 3D clustering measurements once the redshifts for the LRG sample have been measured, and a reference that will aid future steps.

\subsection{Zero point sensitivities}
\label{sec:zp_sensitivity}

Another source of imaging systematics is the zero point (ZP) uncertainty –– the uncertainty of the photometric zero point for each exposure of the imaging data. While depth and seeing can be accurately measured and their effects on targeting can in principle be modeled, the ZP uncertainty is a systematic uncertainty that is mostly due to imperfect modeling of the observing conditions. Thus, the imprint of ZP uncertainties on the sky is difficult, if not impossible, to correct for. We design the LRG target selection to be as insensitive to the ZP uncertainties as possible.


One way to quantify the effects of ZP uncertainties on the LRG targets is to estimate the level of fluctuation in target density caused by the ZP uncertainties. The estimated ZP uncertainties in $g,r,z,W1$ are 3,3,6,1 mmag, respectively \citep{schlegel_dr9}. For the LRG selection, a net change of +10 mmag in g, r, z, W1 causes a change of +0.11\%, +1.40\%, -1.23\%, -2.89\% in target density, respectively. If we assume that the ZP errors in the four bands are all uncorrelated with each other, we can treat the combined effect on the target density as sums of independent Gaussian random variables, and the resulting RMS of the target density is 0.9\%. During Survey Validation, we considered an alternative LRG selection that implements the luminosity cut using $z$ band fiber magnitude and $r-z$ color (see Appendix \ref{sec:sv1}), and this selection has a much larger RMS of ${\sim}\,4\%$ due to the large ZP uncertainty in the $z$ band. (The $z$-band photometric calibration has large uncertainties mainly because the effective $z$-band filter transmission can vary significantly due to telluric water vapor absorption.) Its insensitivity to zero point errors is the main reason that we chose to adopt the WISE-based luminosity cut.

\section{Spectroscopic assessment}
\label{sec:spectro_assessment}


\subsection{Spectroscopic data}
\label{sec:spectro_data}

\begin{figure*}
    \centering
    \includegraphics[width=1.95\columnwidth]{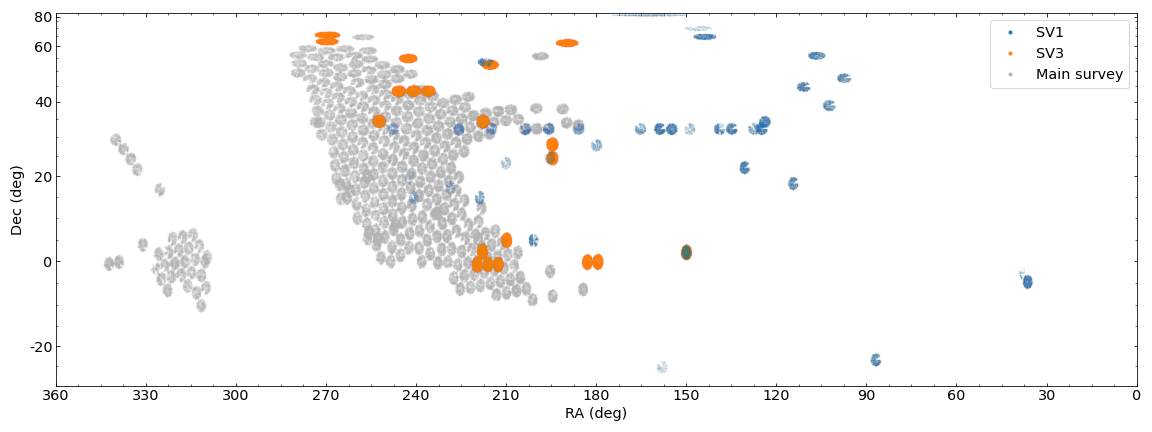}
    \caption{LRGs observed by DESI during Survey Validation and the first two months of the Main Survey. We use this data for evaluating the LRG redshift performance.}
    \label{fig:observed_lrgs}
\end{figure*}

\begin{figure*}
    \centering
    \includegraphics[width=2.1\columnwidth]{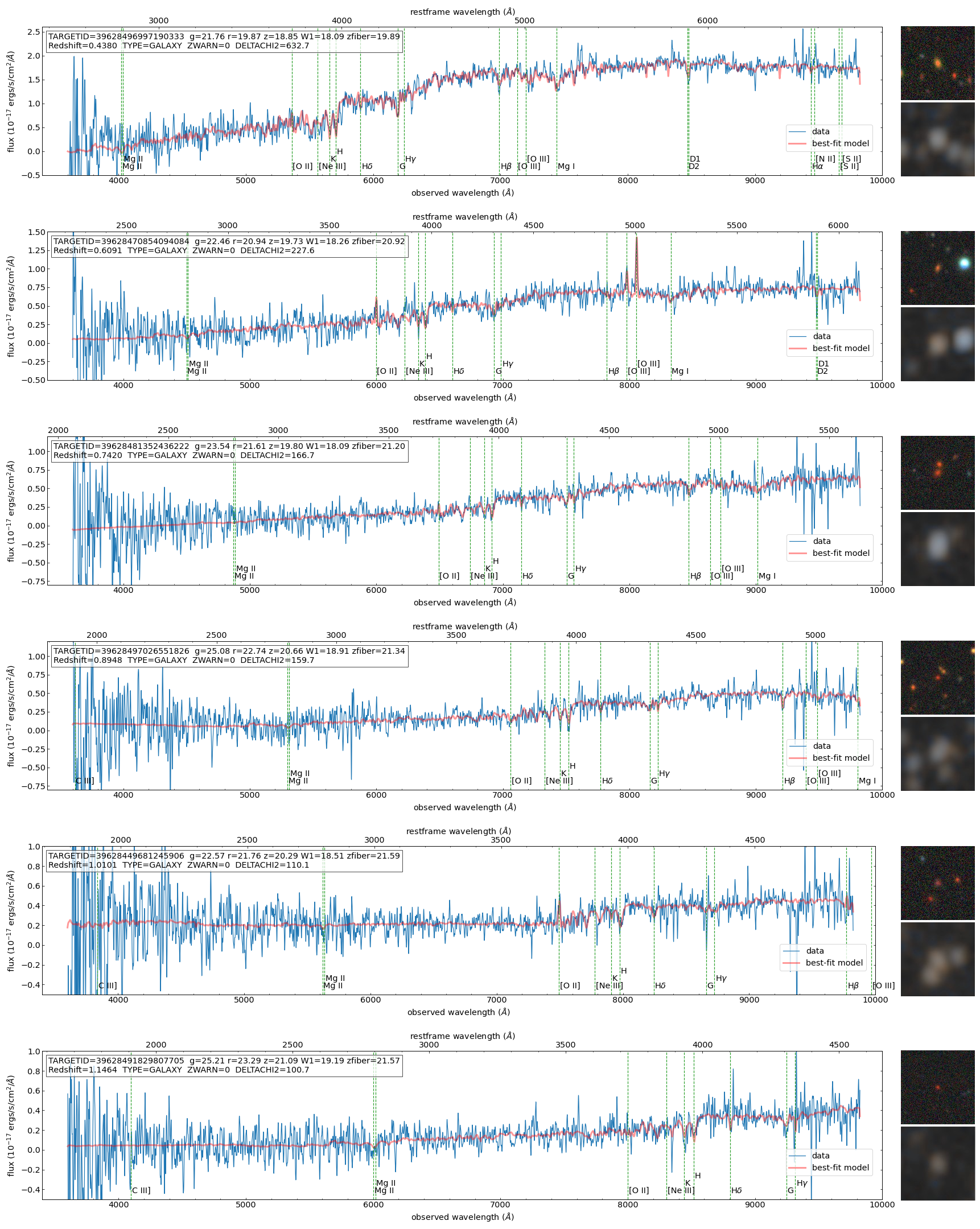}
    \caption{Example spectra and image cutouts of DESI LRGs that were observed in the Main Survey to nominal spectroscopic depth. The observed and model spectra are convolved with a Gaussian kernel with $\sigma=2.4\text{\AA}$ to reduce the noise. The three spectra from the B/R/Z spectrographs are coadded into a single spectrum in the figure. The target ID, $g/r/z/W1$ magnitudes, $z_\mathrm{fiber}$ magnitude, best-fit redshift, best-fit spectral type, ZWARNING flag, and $\Delta \chi^2$ values are listed for each object. Major absorption and emission lines, which are taken from the DESI visual inspection tool \emph{Prospect} (\url{https://github.com/desihub/prospect}), are shown in green dashed lines. The image cutouts are 34\arcsec$\times$34\arcsec composites in $g/r/z$ (top) and $W1/W2$ (bottom).}
    \label{fig:example_spectra}
\end{figure*}

We use spectroscopic data from SV1, SV3, and the first 2 months of the Main Survey. We only select LRG+QSO tiles in SV1 and dark tiles in SV3 and the Main Survey. The sky coverage of the spectroscopic LRGs is shown in Figure \ref{fig:observed_lrgs}. Figure \ref{fig:example_spectra} shows some example spectra of LRGs observed during the Main Survey, and the image cutouts (from the Legacy Surveys Viewer\footnote{\url{https://www.legacysurvey.org/viewer}}).

SV1 has several flavors of coadds, and we use the single-exposure coadds, the nominal (1$\times$) depth coadds, and the cumulative (deep) coadds. LRG targets are assigned the target bit $2^0$ in all three programs. A significant number of brighter LRGs are also observed in the BGS program under very different observing conditions, and we do not include them here. We remove objects affected by instrument issues by requiring that the COADD\_FIBERSTATUS value in the catalogs is equal to 0. We apply the veto mask (\S \ref{sec:masks}) to create a clean sample.

The redshift fitting is done using Redrock \citep{redrock}. It uses 1-D spectra produced by the DESI spectroscopic pipeline \citep{guy_spectro_pipeline} as input. For each spectrum, it computes the best-fit redshift and $\Delta \chi^2$, which is the difference in $\chi^2$ between the best-fit model and the second best-fit model and is an indication of the reliability of the best-fit redshift.

We use a sample of ``true'' redshifts as the reference sample for measuring the catastrophic failure rates. While the redshifts from visual inspection \citep{lan_desi_galaxy_vi} could also be used as true redshifts, they are only available for a few thousand SV LRGs, and a few hundred Main Survey LRGs. Therefore we use the much larger sample of redshifts from deep coadded SV1 spectra as the true redshifts. We require a minimal effective exposure time ($t_\mathrm{eff}$, see its definition in \citealt{guy_spectro_pipeline}) of 3000s, which is 3 times the DESI nominal depth of $t_\mathrm{eff}$=1000s. We validate these deep redshifts by comparing them with the visual inspection redshifts, and we find that the redshifts disagree for less than 0.5\% of the Main Survey LRGs.

For assessing the spectroscopic performance at nominal depth, we require $t_\mathrm{eff}>$800s. And we also require $t_\mathrm{eff}<$1200s for the SV1 coadds (which have a wider range of $t_\mathrm{eff}$ than the Main Survey). To assess if a redshift (e.g., obtained at nominal depth) is correct, we compare it with the redshift of the same object from the deep coadded spectra. If it differs from the deep redshift by more than 1000 km/s, that redshift is considered a ``catastrophic redshift failure''. (A small number of deep redshifts have ZWARN$\neq$0 or $z_\mathrm{redrock}>1.5$, and are likely not reliable. We treat the corresponding nominal-depth redshifts as catastrophic failures regardless of their redshift values.)

Figure \ref{fig:main_dndz} shows the redshift distribution of LRGs observed in the first two months of the Main Survey that covers roughly 2000 deg$^{-2}$. The sample has a roughly constant comoving density of $5\times10^{-4}\ h^3\mathrm{Mpc}^{-3}$ in $0.4<z<0.8$ and a high-z tail that extends beyond $z=1.0$. Figure \ref{fig:main_dndz} excludes a small ($1\%$) fraction of the observed LRG targets that are rejected by the quality cut (see \S\ref{sec:z_quality_cut}); these objects are the faintest LRG targets and are mostly at the high-redshift end of the sample.

\subsection{Spectroscopic classification}
\label{sec:spectro_classification}

Only 0.5\% of the LRG targets are spectroscopically classified by Redrock as stars and 1.7\% as QSOs. The rest are classified as galaxies. Almost all the stellar contaminants are cool stars such as M dwarfs whose strong infrared emissions allow them to pass the stellar-rejection cut. About 0.6\% of the LRG targets are also QSO targets,
and among them, 61\% are spectroscopically classified as QSOs and 2\% as stars. If we exclude QSO targets, the stellar fraction is 0.5\% and the QSO fraction is 1.2\%. A large fraction of the area observed in the first 2 months of the Main Survey is at relatively low Galactic latitude, and the full DESI footprint will include a larger fraction of high-latitude area where the stellar density is lower. Therefore we expect the 0.5\% stellar contamination rate to be an upper bound for the full DESI sample.

Note that in the first two months of the Main Survey observations, the observed LRG targets include a larger fraction (by a factor of $\sim 2$) of objects that are also QSO targets than in the full survey. This is because the overall fiber-assignment completeness is lower at the beginning of the survey and the QSO targets have a higher fiber-assignment priority than the LRGs (see \citealt{raichoor_fiberassign}). We correct for this by downweighting the QSO targets so that the fractions described above are estimates for the final Main Survey sample.



\subsection{Redshift failure rate and quality cut}
\label{sec:z_quality_cut}

The catastrophic redshift failure rate of LRGs observed at nominal depth is $0.7\%$ (110/15379) from comparing with the deep redshifts. This is consistent with what we find from repeat observations: in the overlap between SV3 and the Main Survey, we find that $1.2\%$ (84/7233) of the repeats have different redshifts, which means that the per-object catastrophic failure rate is $0.6\%$ if we assume that the redshift efficiency is the same in SV3 and Main Survey.

\begin{figure}
    \centering
    \includegraphics[width=0.95\columnwidth]{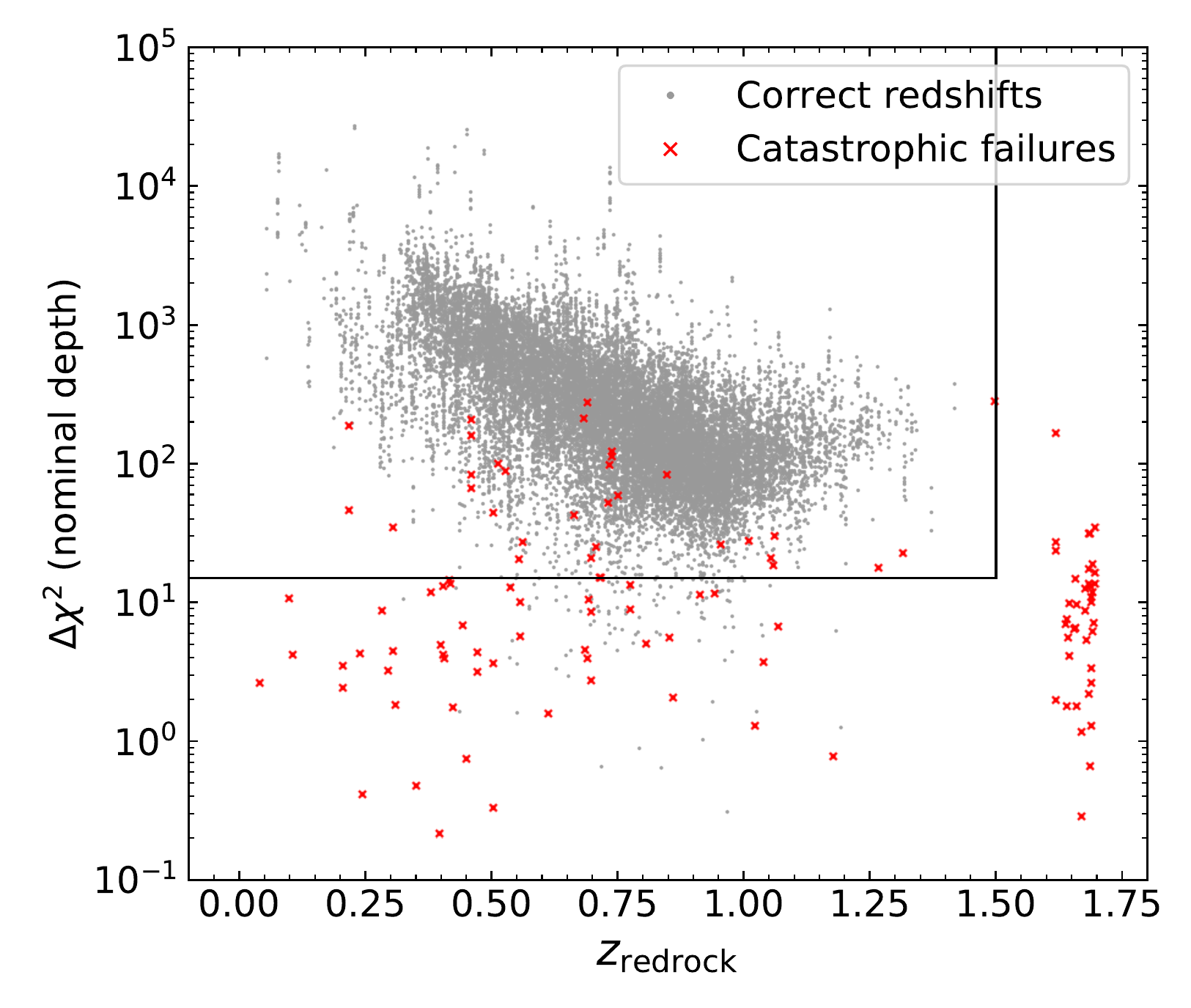}
    \caption{$\Delta \chi^2$ vs redshift (best-fit from Redrock) at nominal depth. The black lines are the redshift quality cut. We distinguish between correct redshifts and catastrophic redshift failures by using redshifts from the deep coadds as truth.}
    \label{fig:dchi2_vs_z}
\end{figure}

To reject incorrect redshifts, we apply the following redshift quality cut (shown in Figure \ref{fig:dchi2_vs_z}): we require $\Delta \chi^2>15$, $z_\mathrm{redrock}<1.5$, and the ZWARNING flag ZWARN=0. The $z_\mathrm{redrock}<1.5$ cut removes the pile-up of catastrophic failures at $z{\sim}\,1.6$. (While it is not entirely clear what causes this pile-up, based on the fact that these objects have mostly noisy and featureless spectra, we speculate that to best match the observed featureless spectra, Redrock finds the best fit at $z>{\sim}\,1.5$ so that the $4000\text{\AA}$ break feature of the PCA templates are red-shifted outside the DESI spectral coverage.) We note that the redshift quality cut is preliminary, and it may change, e.g., as the spectroscopic pipeline and Redrock evolve.

The redshift quality cut removes $1.1\%$ of the LRGs, $43\%$ (82/191) of which are catastrophic failures. The catastrophic failure rate in the accepted (confident) redshifts is $0.2\%$ (28/15188). From visual inspection of the spectra and images, we find that a significant fraction of the catastrophic failures that pass the redshift quality cut are blends (i.e., there is more than one object within the fiber diameter), and both redshift solutions produce reasonable fits to the observed spectra.

Hereafter, we refer to the fraction of objects that meets the quality cut as the ``redshift success rate'' and the fraction that fails the cut as the ``failure/rejection rate'', and we refer to the incorrect redshifts (as determined by comparing with deep or repeat spectra) as ``\textit{catastrophic} failures''.

In the above assessments, we have excluded QSO targets and objects spectroscopically classified as QSOs, because they have much higher redshift failure rates and are a very different population than the ``normal'' LRGs. The objects that are targeted or classified as QSOs have approximately 10 times higher catastrophic failure rates than the rest of the LRG targets, and they are also about 10 times more likely to fail the redshift quality cut. It should be noted that the distinction between a ``galaxy'' and ``QSO'' is not always sharply defined (e.g., Redrock may classify some ``galaxies'' with AGN features as ``QSOs'' but not others), and more careful consideration may be needed for the selection of the DESI clustering sample.


\subsection{Depth and magnitude dependence}
\label{sec:depth_and_mag_dependence}

The spectroscopic redshifts of the LRGs are primarily based on absorption lines and the $4000\text{\AA}$ break, and sufficient spectroscopic S/N is critical for confident redshift estimation. The S/N mainly depends on two factors: the source brightness and the spectroscopic depth ($t_\mathrm{eff}$). Here we investigate how the two factors affect the LRG redshift failure rate. (The redshift failure rate also depends on other factors such as the strength of the absorption and emission lines and prominence of the $4000\,\text{\AA}$ break, but they do not vary significantly within the LRG sample and are thus less important factors for the redshift determination.)

\begin{figure}
    \centering
    \includegraphics[width=0.95\columnwidth]{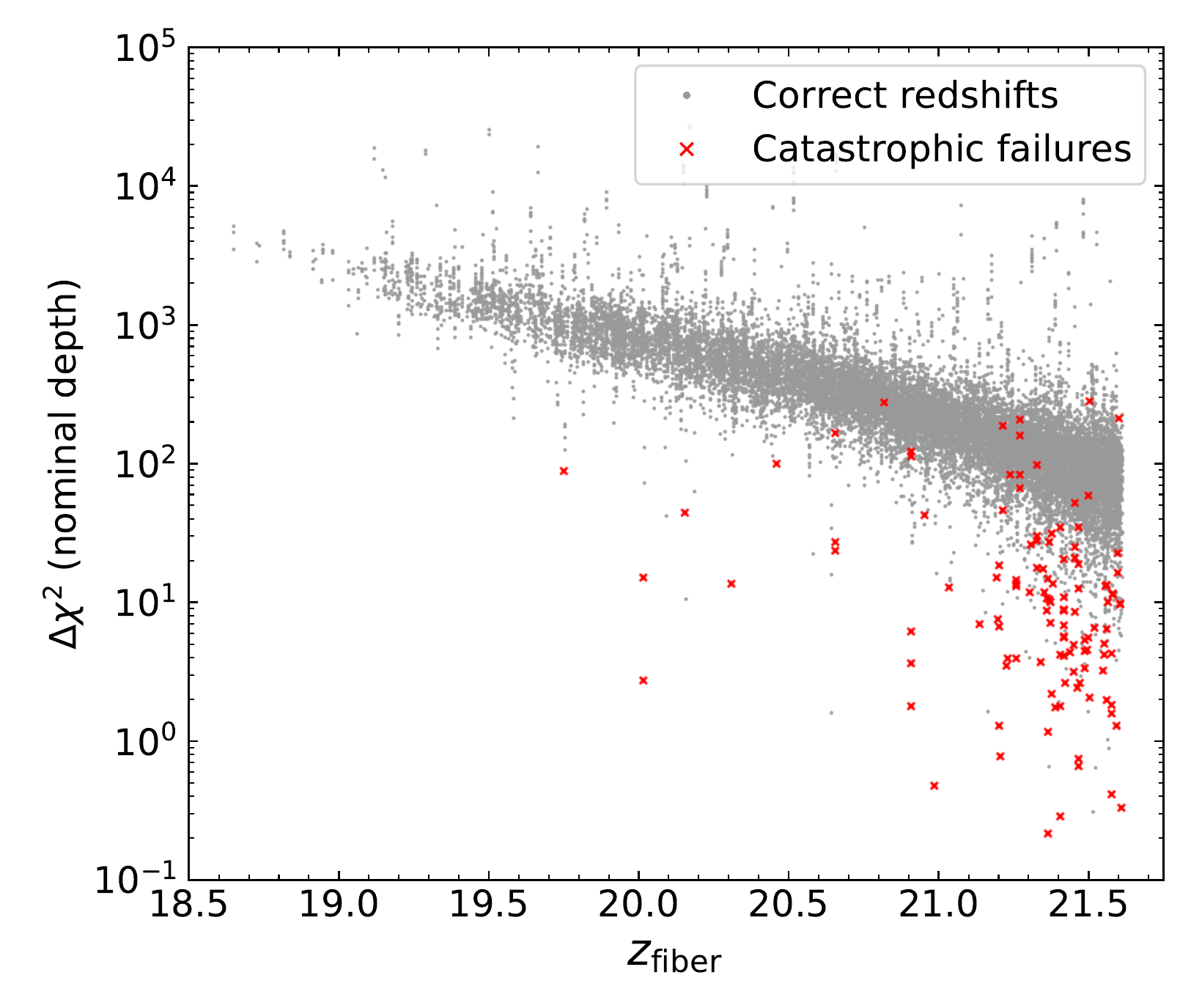}
    \caption{Similar to Figure \ref{fig:dchi2_vs_z} but with x-axis replaced by $z$-band fiber magnitude.}
    \label{fig:dchi2_vs_zfiber}
\end{figure}

\begin{figure}
    \centering
    \includegraphics[width=0.95\columnwidth]{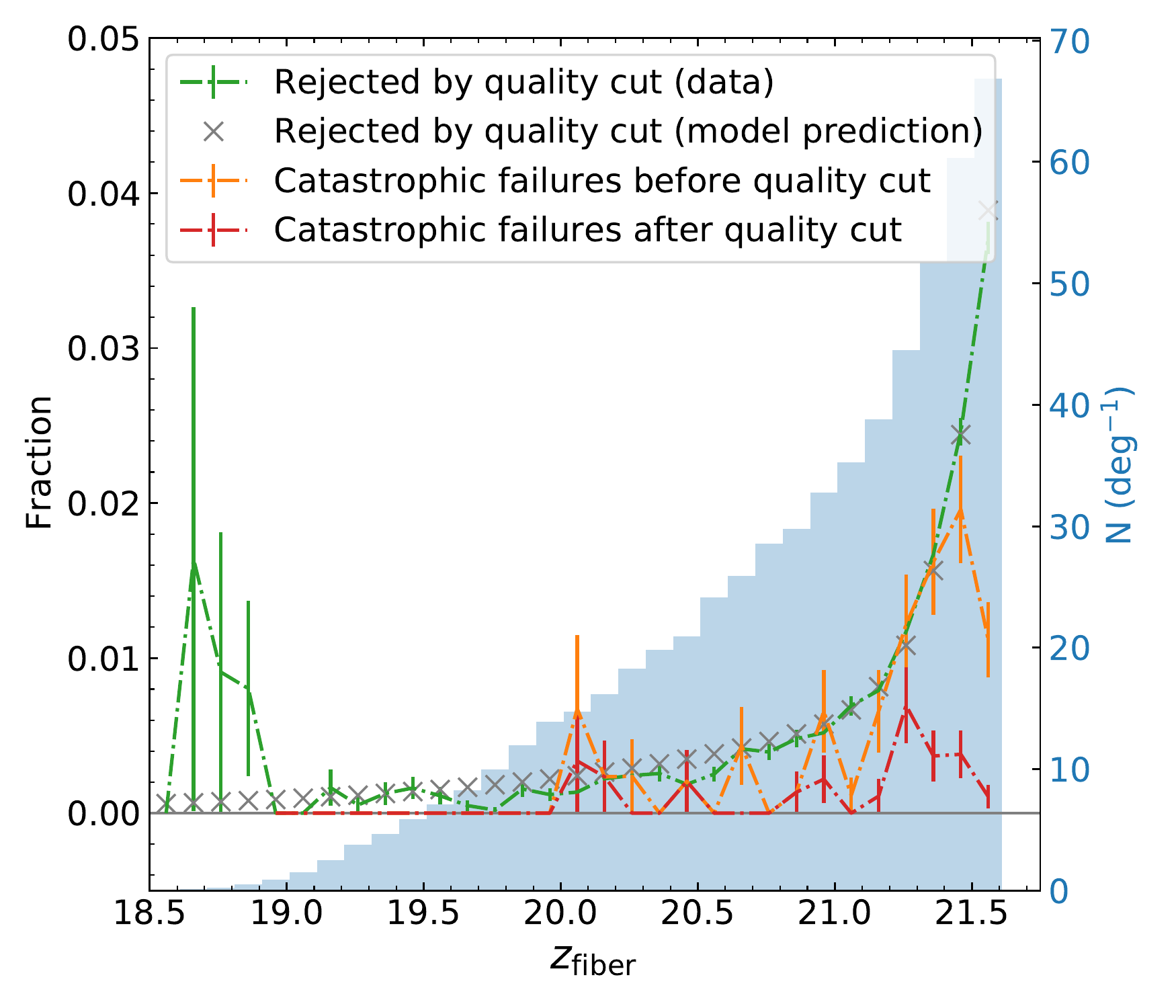}
    \caption{Catastrophic failure rates and rejection rates as a function of $z$-band fiber magnitude for the nominal exposures. (Error bars are not shown for fractions equal to zero.) The gray crosses are the predicted rejection rates based on $z_\mathrm{fiber}$ and $t_\mathrm{eff}$ (see \S\ref{sec:depth_and_mag_dependence}). The histogram shows the $z_\mathrm{fiber}$ distribution of the LRG sample. We restrict to coadds with 800s $< t_\mathrm{eff} <$ 1200s.}
    \label{fig:failure_rate_vs_zfiber}
\end{figure}

The $z$-band fiber magnitude ($z_\mathrm{fiber}$) is very well correlated with $\Delta \chi^2$ and catastrophic redshift failures, as shown in Figure \ref{fig:dchi2_vs_zfiber}, and the correlation is much stronger than the fiber magnitudes in the $g$ and $r$ bands. Therefore we use $z_\mathrm{fiber}$ as the parameter for source brightness. In Figure \ref{fig:failure_rate_vs_zfiber} we show the catastrophic redshift failure rate and rejection rate as a function of $z_\mathrm{fiber}$ at nominal depth. The error bars show the uncertainty for a binomial distribution: $\sigma_p=\sqrt{N p (1-p)}/N$ where $N$ is the total number of objects and $p$ is the failure/rejection rate. At the $z_\mathrm{fiber}$ limit, the LRGs have the highest catastrophic failure rate of ${\sim}\,2\%$ and rejection rate of ${\sim}\,4\%$. The catastrophic rate after applying the redshift quality cut is much less than $1\%$ at the $z_\mathrm{fiber}$ limit.

\begin{figure}
    \centering
    \includegraphics[width=0.95\columnwidth]{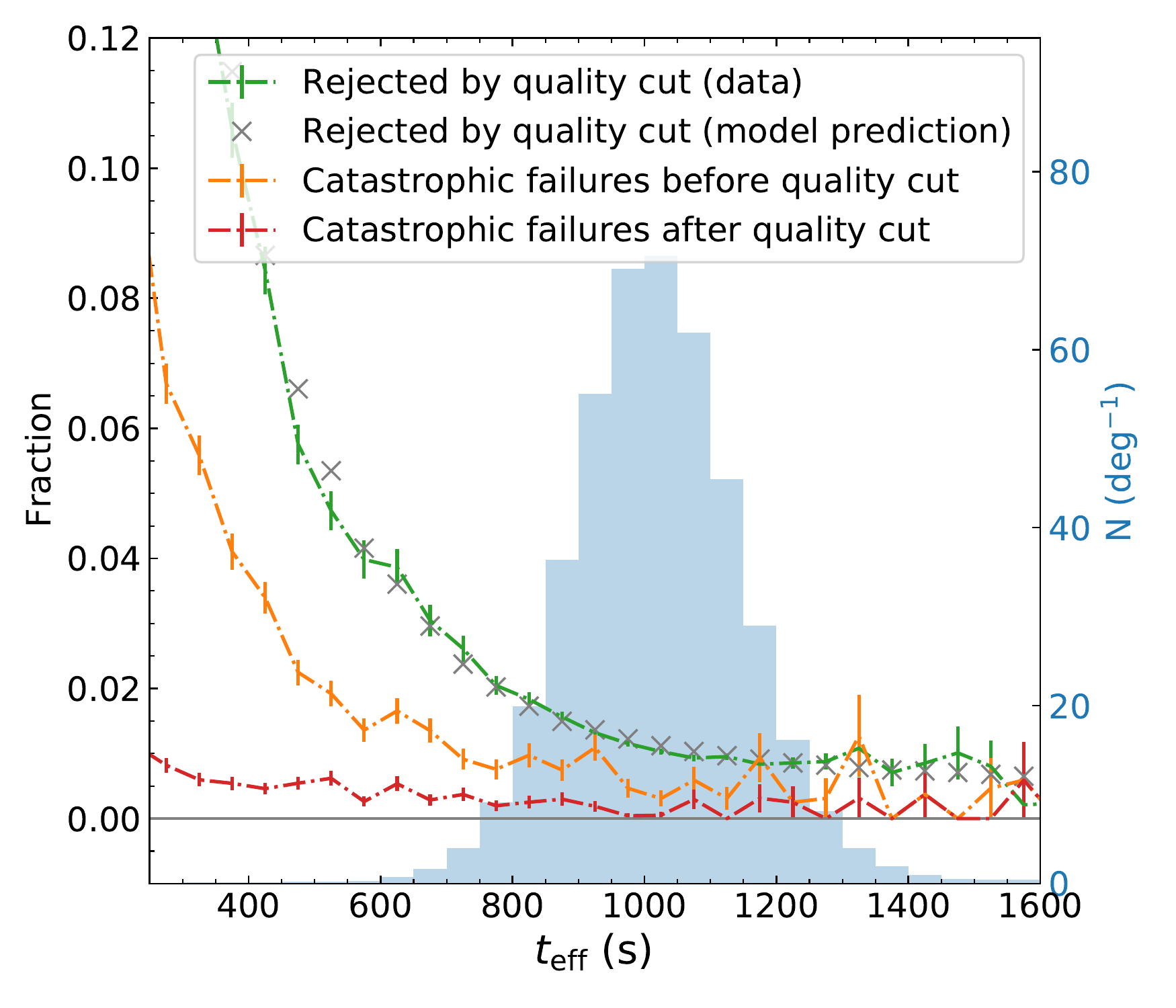}
    \caption{Similar to Figure \ref{fig:failure_rate_vs_zfiber}, but with the effective exposure time $t_\mathrm{eff}$ in the x-axis (and with the restriction on $t_\mathrm{eff}$ removed). The histogram shows the $t_\mathrm{eff}$ distribution of LRGs in the Main Survey.}
    \label{fig:failure_rate_vs_efftime}
\end{figure}

Figure \ref{fig:failure_rate_vs_efftime} shows the catastrophic redshift failure rate and the rejection rate as a function of $t_\mathrm{eff}$. The rejection rate flattens out at above $t_\mathrm{eff}={\sim}\,$1000s. And while the catastrophic failure rate and rejection rate increase significantly at $t_\mathrm{eff}<$800s, the catastrophic failure rate of the accepted redshifts remains well below $1\%$.

For clustering measurements, it is important to correct for redshift incompleteness caused by targeting and observational factors. Here we use the following function of the effective exposure time and $z$-band fiber flux to predict the LRG redshift failure/rejection probability:
\begin{equation}
P_\mathrm{fail}(t_\mathrm{eff}, f_{z,\mathrm{fiber}}) = \exp(a_0 S + a_1) + a_2/(f_{z,\mathrm{fiber}}/1\,\mathrm{nanomaggy}) ,
\label{eqn:failure_rate}
\end{equation}
where $S \equiv (f_{z,\mathrm{fiber}}/1\,\mathrm{nanomaggy})\sqrt{t_\mathrm{eff}/1\,\mathrm{sec}}$ is (approximately) proportional to the spectroscopic S/N when the sky is much brighter than the source (which is true for the fainter LRGs), $f_{z,\mathrm{fiber}}$ is the $z$-band fiber flux, and $a_0$, $a_1$, $a_2$ are constant coefficients.

The exponential term is motivated by the observation that the redshift failure rate decreases exponentially with $S$. At brighter magnitudes and in deeper exposures (where the per-object failure rate becomes less than ${\sim}\,1\%$) the exponential term approaches zero faster than the observed failure rate, so we add the second term $a_2/f_{z,\mathrm{fiber}}$ to account for the higher observed redshift failure rate. From visual inspection, we find that many of the redshift failures at brighter magnitudes are blends or have problematic spectra due to instrument issues.

The best-fit coefficients are found by minimizing
$\sum_i (P_{\mathrm{fail}, i} - Q_i)^2$, 
where $P_{\mathrm{fail},i}$ is the predicted failure probability of the $i$-th object, and $Q_i=0$ if the object passes the quality cut and $Q_i=1$ if it fails the cut. For the fitting we use Main LRGs observed in SV1 and the first 2 months of the Main Survey, and SV3 LRGs (which are ${\sim}\,$0.1 magnitude fainter than the Main LRGs). To best match the redshift failure rates of the Main Survey, we only use LRGs with $500s<t_\mathrm{eff}<2000s$; this prevents the large number of redshift failures at very low $t_\mathrm{eff}$ (mainly from SV1) from dominating the fit. The best-fit coefficients are $a_0=-0.0911$, $a_1=3.34$, $a_2=0.0228$.

\begin{figure*}
    \centering
    \includegraphics[width=2.\columnwidth]{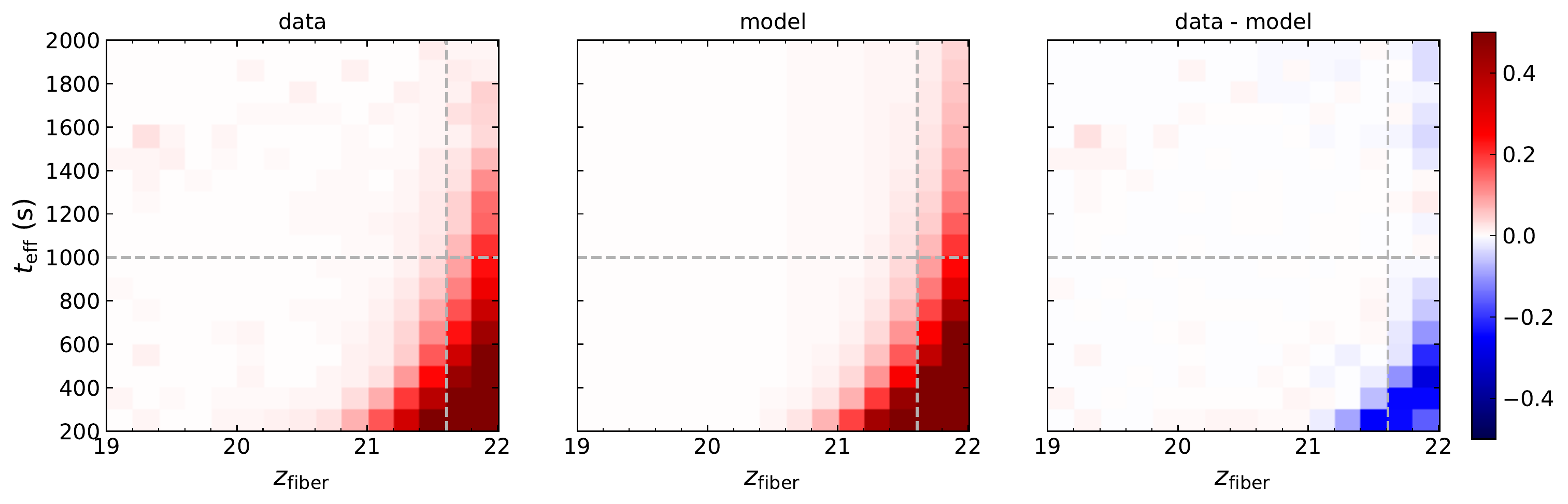}
    \caption{\textit{Left}: Redshift failure rate in bins of spectroscopic depth ($t_\mathrm{eff}$) and fiber magnitude ($z_\mathrm{fiber}$). The horizontal dashed line marks the nominal depth of 1000s. The vertical line marks the magnitude limit ($z_\mathrm{fiber}<21.6$) of the Main LRGs. At $z_\mathrm{fiber}<21.6$ the failure rate is computed for the combined sample of Main and SV3 LRGs, and at $z_\mathrm{fiber}\geq21.6$ SV1 LRGs are added. \textit{Middle}: the redshift failure rate from the model prediction using eqn. \ref{eqn:failure_rate}. \textit{Right}: the residual, i.e., the measured failure rate subtracted by the predicted failure rate. The model fitting is done using Main and SV3 LRGs with $t_\mathrm{eff}>500$, which is why the prediction is less accurate at lower $t_\mathrm{eff}$ and in the faintest magnitude bin.}
    \label{fig:failure_rate_in_2d}
\end{figure*}

The gray crosses in Figures \ref{fig:failure_rate_vs_zfiber} and \ref{fig:failure_rate_vs_efftime} are the predicted failure rates, and they match the observed failure rates (green error bars) very well. Figure \ref{fig:failure_rate_in_2d} shows the observed and predicted dependence of the LRG redshift failure rate on both $z_\mathrm{fiber}$ and $t_\mathrm{eff}$. The residuals are negligible in the range of $z_\mathrm{fiber}$ and $t_\mathrm{eff}$ relevant for LRGs in the Main Survey.


\subsection{Fiber-to-fiber variation in failure rate}
\begin{figure*}
\gridline{\fig{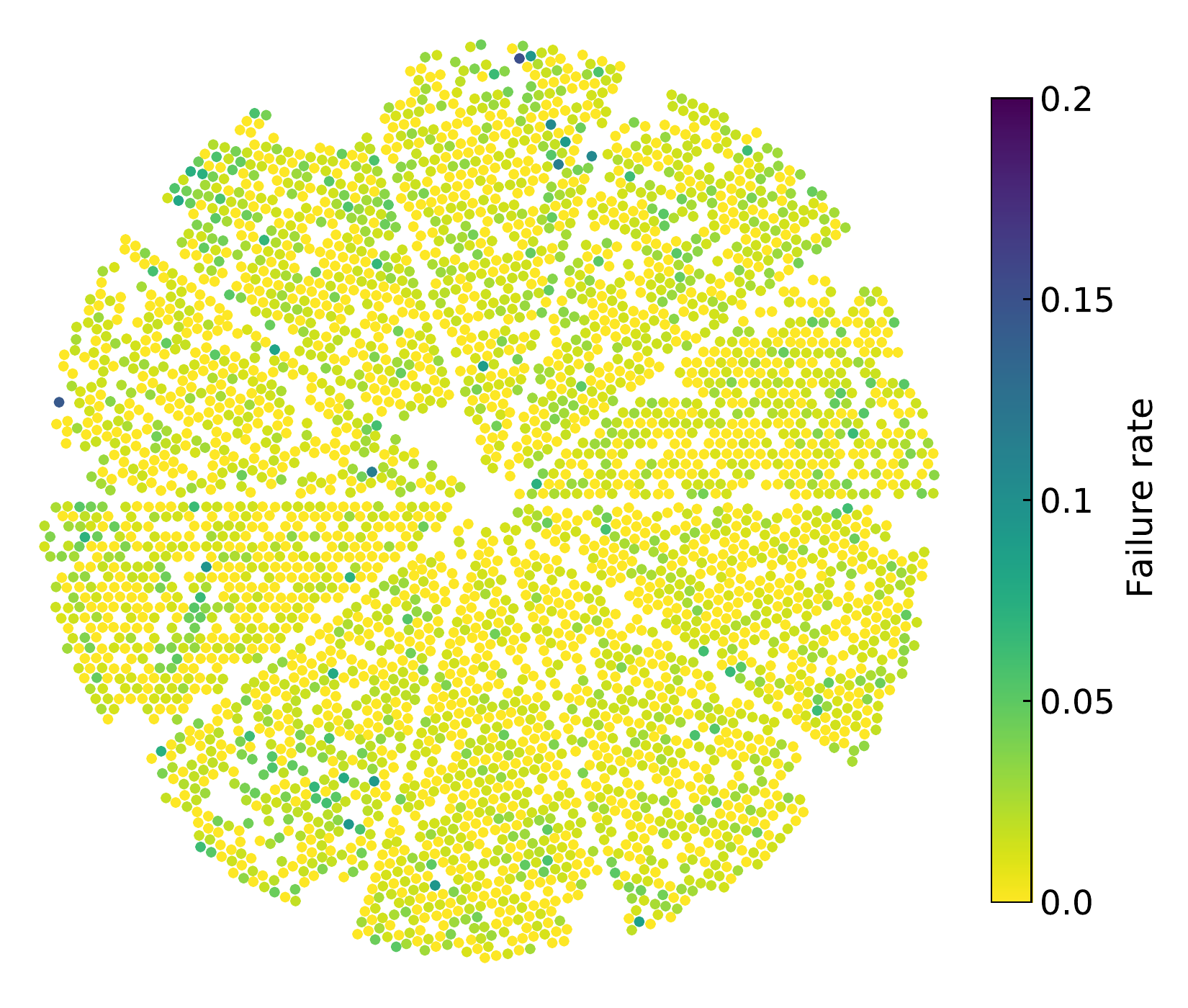}{1.05\columnwidth}{}
          \fig{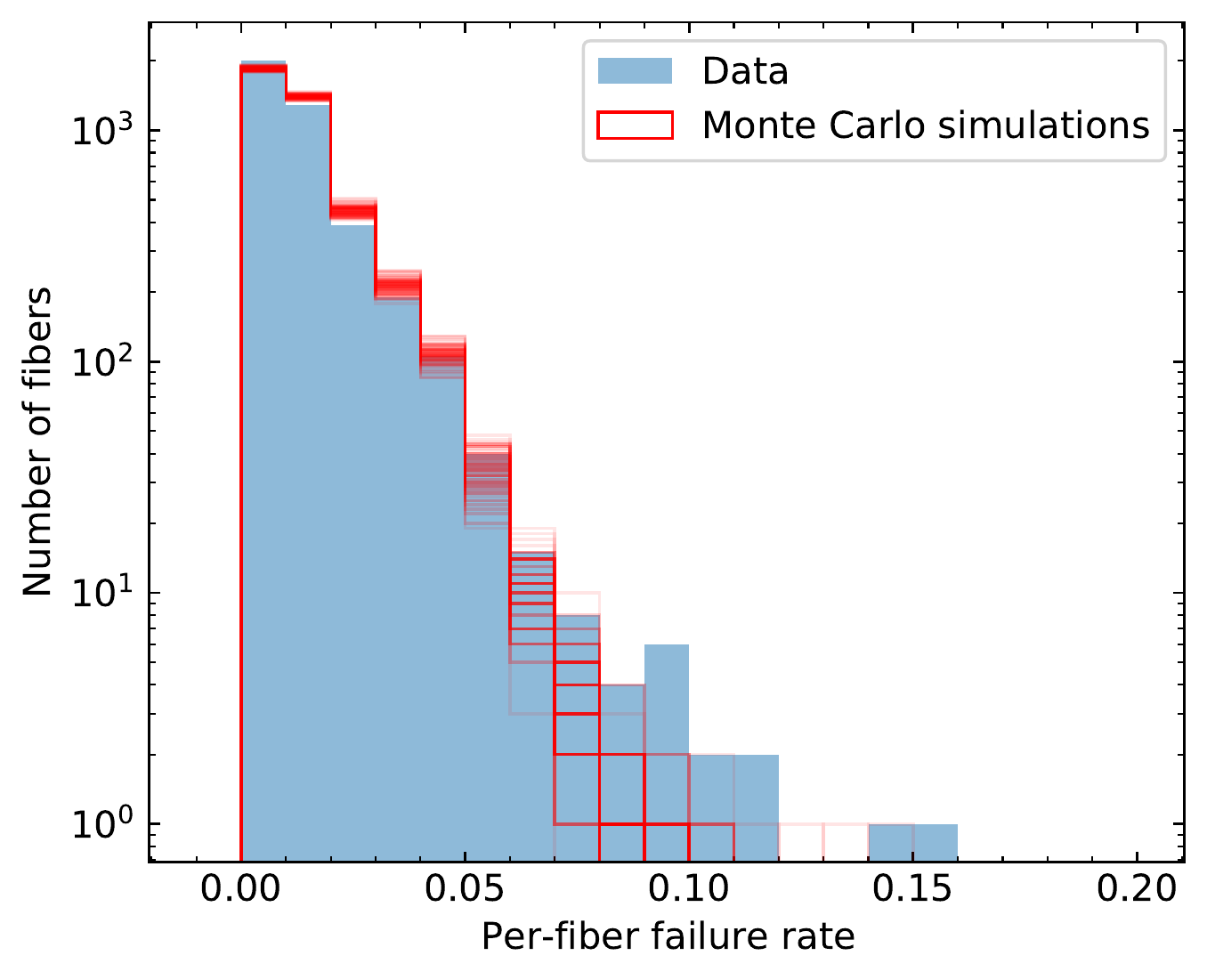}{0.95\columnwidth}{}
          }
\caption{\textit{Left}: Per-fiber LRG redshift failure rate. Each point represents a fiber on the DESI focal plane with the colors indicating its average LRG failure rate during the first 2 months of the Main Survey. Only fibers with $>$40 LRG observations are plotted, and the median number of LRG observations by each fiber is 70. Since the average failure rate is ${\sim}\,1\%$, most fibers have either 0 or 1 redshift failure, and much of the variation in this figure is simply noise. \textit{Right}: The distribution of the per-fiber LRG redshift failure rates and the simulated distributions from 100 Monte Carlo simulations. In the simulations, the redshift failure probability of each object is determined by its $z_\mathrm{fiber}$ and $t_\mathrm{eff}$ via eqn. \ref{eqn:failure_rate}. The fact that the measured distribution matches the simulations (for all except a handful of fibers) suggests that the spectroscopic efficiency is very uniform across the fibers.
\label{fig:per_fiber_failure_rate}}
\end{figure*}

For the DESI clustering analysis, it is important that variations in the spectroscopic efficiency of each fiber do not imprint on the measured galaxy densities. E.g., \citet{ross_clustering_2012} found that the redshift efficiency of the BOSS CMASS sample varies with the fiber location on the focal plane.  To quantify the per-fiber efficiency, we compute the average LRG redshift failure rate for each fiber using observations during the Main Survey. The per-fiber LRG failure rate is shown in Figure \ref{fig:per_fiber_failure_rate} (left panel), and we do not see clear patterns in the fiber efficiency given the statistical uncertainties. To more rigorously assess if the observed failure rates are consistent with uniform fiber efficiency, we perform Monte Carlo simulations with the per-object failure probability given by the model described in \S\ref{sec:depth_and_mag_dependence}. We find that the observed distribution of the measured per-fiber failure rates is mostly consistent with the simulated distributions for all except a few fibers, as shown in Figure \ref{fig:per_fiber_failure_rate} (right panel), indicating very uniform spectroscopic efficiencies for almost all the fibers. We will revisit the fiber-to-fiber uniformity when more data becomes available.

\section{Summary}
\label{sec:summary}

To achieve the required accuracy on cosmological measurements for DESI, it is critical that the sample selection and spectroscopic observations have minimal and well-understood systematics. With this in mind, the DESI LRG sample is designed to be robust against variations in imaging properties and zero point uncertainties and to achieve a high redshift success rate and low stellar contamination rate. The high stellar mass completeness of the sample also ensures high large-scale bias and should facilitate modeling of the galaxy-halo connection. In addition to the already robust target selection, we developed veto masks specifically optimized for the LRG targets to produce a clean sample suitable for clustering analysis. We also created a simple model that can accurately predict (and thus correct for) the per-object redshift failure rate based on the object's brightness and spectroscopic depth.

The LRG target catalogs are publicly available\footnote{\url{https://data.desi.lbl.gov/public/ets/target/catalogs/dr9/1.1.1/targets/main/resolve/dark/}} (see \citealt{myers_target_pipeline} for details).
The data points in all figures are publicly available at Zenodo: \dataset[doi:10.5281/zenodo.6987401]{https://doi.org/10.5281/zenodo.6987401}.





\section*{acknowledgments}

This research is supported by the Director, Office of Science, Office of High Energy Physics of the U.S. Department of Energy under Contract No. DE–AC02–05CH11231, and by the National Energy Research Scientific Computing Center, a DOE Office of Science User Facility under the same contract; additional support for DESI is provided by the U.S. National Science Foundation, Division of Astronomical Sciences under Contract No. AST-0950945 to the NSF's National Optical-Infrared Astronomy Research Laboratory; the Science and Technologies Facilities Council of the United Kingdom; the Gordon and Betty Moore Foundation; the Heising-Simons Foundation; the French Alternative Energies and Atomic Energy Commission (CEA); the National Council of Science and Technology of Mexico (CONACYT); the Ministry of Science and Innovation of Spain (MICINN), and by the DESI Member Institutions: \url{https://www.desi.lbl.gov/collaborating-institutions}.

The DESI Legacy Imaging Surveys consist of three individual and complementary projects: the Dark Energy Camera Legacy Survey (DECaLS), the Beijing-Arizona Sky Survey (BASS), and the Mayall z-band Legacy Survey (MzLS). DECaLS, BASS and MzLS together include data obtained, respectively, at the Blanco telescope, Cerro Tololo Inter-American Observatory, NSF's NOIRLab; the Bok telescope, Steward Observatory, University of Arizona; and the Mayall telescope, Kitt Peak National Observatory, NOIRLab. NOIRLab is operated by the Association of Universities for Research in Astronomy (AURA) under a cooperative agreement with the National Science Foundation. Pipeline processing and analyses of the data were supported by NOIRLab and the Lawrence Berkeley National Laboratory. Legacy Surveys also uses data products from the Near-Earth Object Wide-field Infrared Survey Explorer (NEOWISE), a project of the Jet Propulsion Laboratory/California Institute of Technology, funded by the National Aeronautics and Space Administration. Legacy Surveys was supported by: the Director, Office of Science, Office of High Energy Physics of the U.S. Department of Energy; the National Energy Research Scientific Computing Center, a DOE Office of Science User Facility; the U.S. National Science Foundation, Division of Astronomical Sciences; the National Astronomical Observatories of China, the Chinese Academy of Sciences and the Chinese National Natural Science Foundation. LBNL is managed by the Regents of the University of California under contract to the U.S. Department of Energy. The complete acknowledgments can be found at \url{https://www.legacysurvey.org/}.

The authors are honored to be permitted to conduct scientific research on Iolkam Du'ag (Kitt Peak), a mountain with particular significance to the Tohono O'odham Nation.

BASS is a key project of the Telescope Access Program (TAP), which has been funded by the National Astronomical Observatories of China, the Chinese Academy of Sciences (the Strategic Priority Research Program ``The Emergence of Cosmological Structures'' Grant \#XDB09000000), and the Special Fund for Astronomy from the Ministry of Finance. The BASS is also supported by the External Cooperation Program of Chinese Academy of Sciences (Grant \#114A11KYSB20160057), and Chinese National Natural Science Foundation (Grant \#12120101003, \#11433005).

ADM was supported by the U.S. Department of Energy, Office of Science, Office of High Energy Physics, under Award Number DE-SC0019022.


\software{Astropy \citep{astropy:2013, astropy:2018}, HEALPix/healpy \citep{healpix, healpy} Matplotlib \citep{matplotlib}, Numpy \citep{numpy}, scikit-learn \citep{scikit-learn}, Scipy \citep{scipy}}


\appendix

\section{SV1 selection and considerations for the final selection}
\label{sec:sv1}

Here we describe the LRG sample observed during Survey Validation (or ``SV1'') before the Main Survey operations. We also discuss the reasoning behind the choices for the Main selection based on data from SV1. See \citet{dawson_sv_overview} for an overview of the Survey Validation program.

\begin{figure*}
    \centering
    \includegraphics[width=2.05\columnwidth]{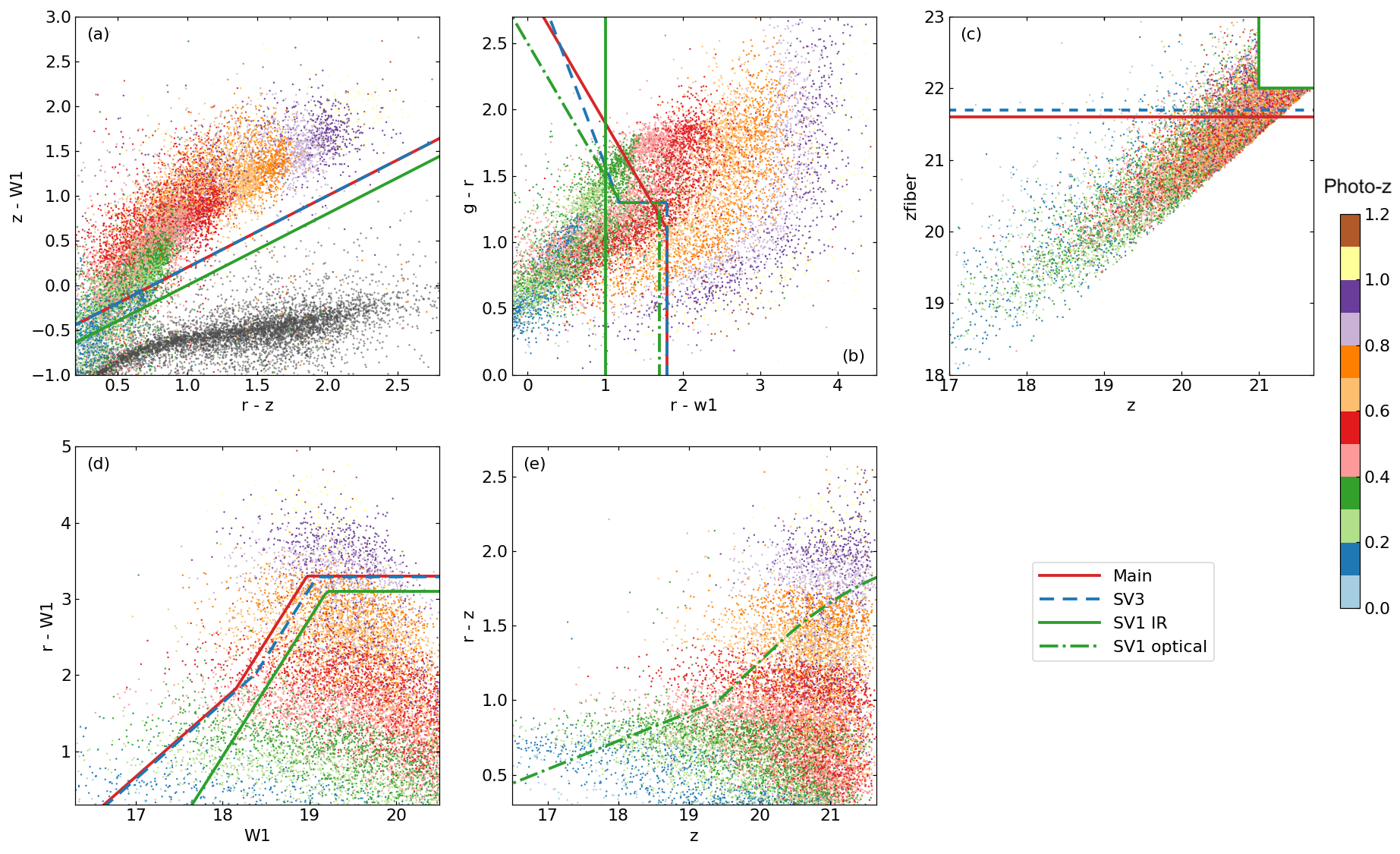}
    \caption{LRG selection cuts for SV1 (solid green and dashed-dotted green lines), SV3 (dashed blue lines), and the Main Survey (solid red lines). The SV1 selection includes two sub-selections, the IR selection (solid green) and optical selection (dashed-dotted green); see text. The cuts shown here are for the South and are slightly different from the cuts in the North.
    The points are color-coded by their PRLS photometric redshifts \citep{zhou_clustering_2021} and the gray points in panel (a) show the stellar locus. The cuts in panels (a)-(d) serve similar purposes to those in Figure \ref{fig:main_lrg_selection}. Panel (e) shows the sliding color-magnitude cut for optical selection that was explored in the SV but not implemented in the Main selection.}
    \label{fig:sv_lrg_selection}
\end{figure*}

\begin{figure}
    \centering
    \includegraphics[width=0.95\columnwidth]{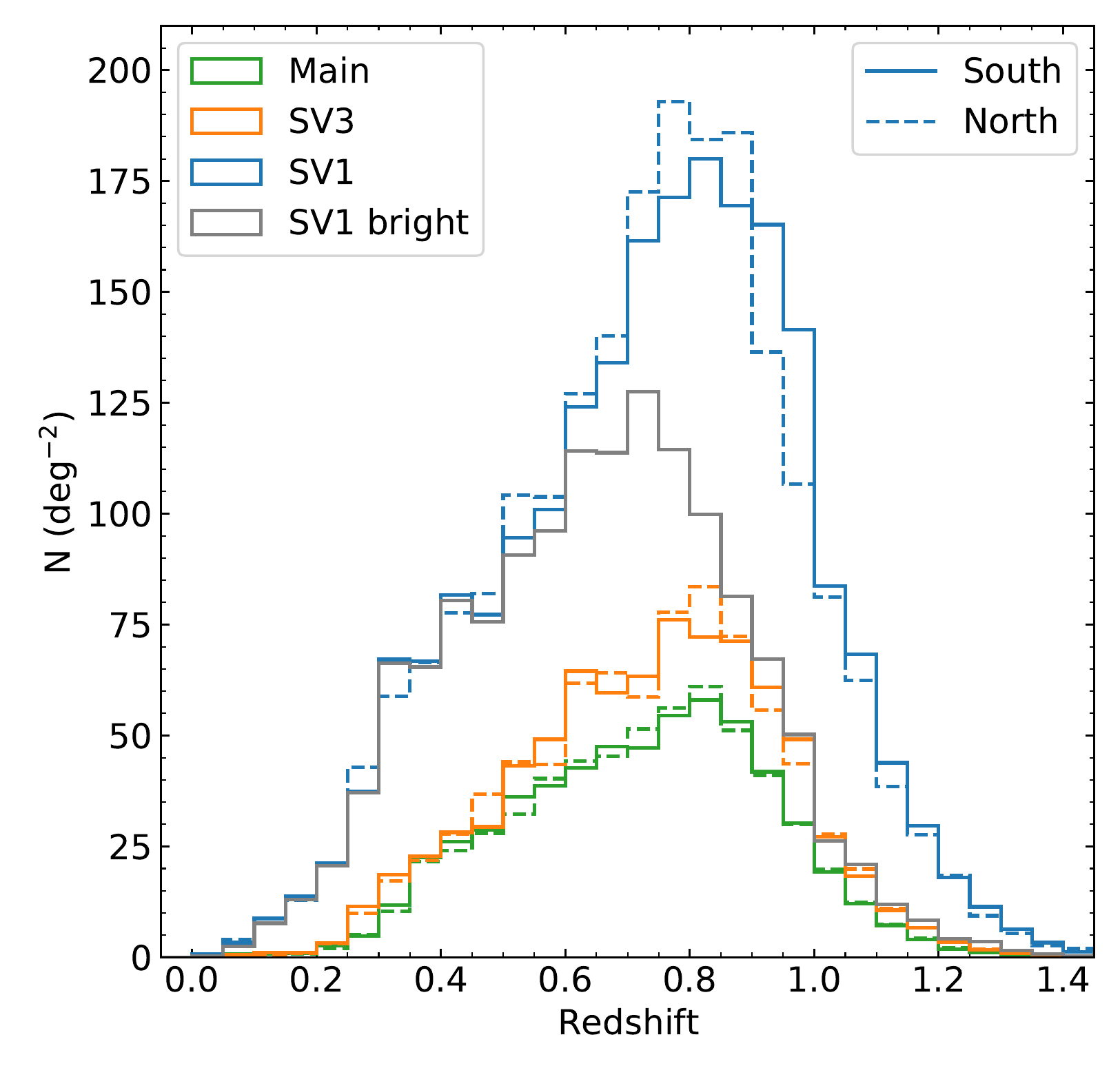}
    \caption{Redshift distributions of the various DESI LRG samples. The ``SV1 bright'' sample (gray, showing South only) is the subset of SV1 LRGs that have the same faint limit as the Main LRGs ($z_\mathrm{fiber}<21.6$) and a very similar redshift success rate as the Main LRGs.}
    \label{fig:all_lrgs_dndz}
\end{figure}

The SV1 selection cuts are listed in Tables \ref{tab:sv1_ir_cuts} (``IR'' selection) and Table \ref{tab:sv1_opt_cuts} (``optical'' selection); an object that meet either selection is selected. The SV1 cuts are shown in Figure \ref{fig:sv_lrg_selection}. The SV1 selection was designed as a parent selection within which we could explore different selection cuts for the final LRG sample, and for this reason, its selection boundaries generally extend beyond the final selection. The SV1 observation is also significantly deeper than the Main Survey, and this allows us to create deeper coadds for assessing the redshift performance. The SV1 redshift distribution is shown in Figure \ref{fig:all_lrgs_dndz}. The target density of the SV1 sample is ${\sim}\,2100$ deg$^{-2}$. The SV1 program observed 46k unique SV1 LRG targets (excluding SV1 LRGs observed in BGS tiles).

\begin{table*}
    \centering
    \caption{SV1 IR selection cuts.}
    \label{tab:sv1_ir_cuts}
    \begin{tabular}{ll}
    \hline
    \hline
    Cuts & Comment \\
    \hline
    \multicolumn{2}{c}{South}\\
    $(z_\mathrm{fiber} < 22.0)  \ \mathrm{OR}\  (z < 21.0)$ & Faint limit\\
    $z-W1 > 0.8 \times (r-z) - 0.8$ & Stellar rejection\\
    $r-W1 > 1.0$ & Remove low-z galaxies\\
    $(r-W1 > (W1-17.48) \times 1.8) \ \mathrm{OR}\ (r-W1 > 3.1)$ & Luminosity cut\\
    \hline
    \multicolumn{2}{c}{North}\\
    $(z_\mathrm{fiber} < 22.0)  \ \mathrm{OR}\  (z < 21.0)$ & Faint limit\\
    $z-W1 > 0.8 \times (r-z) - 0.8$ & Stellar rejection\\
    $r-W1 > 1.03$ & Remove low-z galaxies\\
    $(r-W1 > (W1-17.44) \times 1.8) \ \mathrm{OR}\ (r-W1 > 3.1)$ & Luminosity cut\\
    \hline
    \end{tabular}
\end{table*}

\begin{table*}
    \centering
    \caption{SV1 optical selection cuts.}
    \label{tab:sv1_opt_cuts}
    \begin{tabular}{ll}
    \hline
    \hline
    Cuts & Comment \\
    \hline
    \multicolumn{2}{c}{South}\\
    $(z_\mathrm{fiber} < 22.0)  \ \mathrm{OR}\  (z < 21.0)$ & Faint limit\\
    $z-W1 > 0.8 \times (r-z) - 0.8$ & Stellar rejection\\
    $((g-W1 > 2.5) \ \mathrm{AND}\  (g-r > 1.3)) \ \mathrm{OR}\ (r-W1 > 1.7)$ & Remove low-z galaxies\\
    $((z < 20.2) \ \mathrm{AND}\ (r-z > (z-17.20) \times 0.45)$ 
    $\mathrm{AND}\ (r-z > (z-14.17) \times 0.19))$ & \multirow{2}{9em}{Luminosity cut}\\
    $\mathrm{OR}\ ((z >= 20.2)$
    $\mathrm{AND}\ (((z-23.18) / 1.3)^2 + (r-z+2.5)^2 > 4.48^2))$ \\
    \hline
    \multicolumn{2}{c}{North}\\
    $(z_\mathrm{fiber} < 22.0)  \ \mathrm{OR}\  (z < 21.0)$ & Faint limit\\
    $z-W1 > 0.8 \times (r-z) - 0.8$ & Stellar rejection\\
    $((g-W1 > 2.57) \ \mathrm{AND}\ (g-r > 1.35)) \ \mathrm{OR}\ (r-W1 > 1.75)$ & Remove low-z galaxies\\
    $((z < 20.2) \ \mathrm{AND}\ (r-z > (z-17.17) \times 0.45)$
    $\mathrm{AND}\ (r-z > (z-14.14) \times 0.19))$ & \multirow{2}{9em}{Luminosity cut}\\
    $\mathrm{OR}\ ((z >= 20.2)$
    $\mathrm{AND}\ (((z-23.15) / 1.3)^2 + (r-z+2.5)^2 > 4.48^2))$ \\
    \hline
    \end{tabular}
\end{table*}

The SV1 sample is selected from two selections: an ``IR'' selection and an ``optical'' selection, with a modified version of the former being chosen as the final LRG selection for the Main Survey. Slightly different versions of the ``optical'' selection were presented in \citet{zhou_preliminary_2020} and \citet{zhou_clustering_2021}. The main difference between the two selections is in the sliding color-magnitude cut (Figure \ref{fig:sv_lrg_selection} panels (d) and (e)). Recall that this cut serves as the luminosity cut and shapes the redshift distribution. The IR selection implements this cut in $r-W1$ vs $W1$, whereas the optical selection does it in $r-z$ vs $z$. Both cuts can be tuned to yield the desired redshift distribution, and both yield similar redshift success rates. The optical selection yields slightly higher stellar mass completeness than the IR selection and the optical selection also has a slightly higher bias (by ${\sim}\,5\%$ based on angular correlation amplitudes at intermediate scales). The main advantage of the IR selection is its robustness to imaging systematics. As we discussed in \S\ref{sec:zp_sensitivity}, the WISE $W1$ photometry is much better calibrated than the ground-based $z$-band, and as a result, the WISE-based IR selection is expected to be more uniform than the $z$ band-based optical selection. While the IR selection may be more sensitive to the effects of blending in WISE (due to its much larger PSF than the optical bands), we did not find any clear evidence of systematics introduced by WISE blending (apart from contamination by bright stars in WISE that can be removed by the veto mask).

\begin{figure}
    \centering
    \includegraphics[width=0.99\columnwidth]{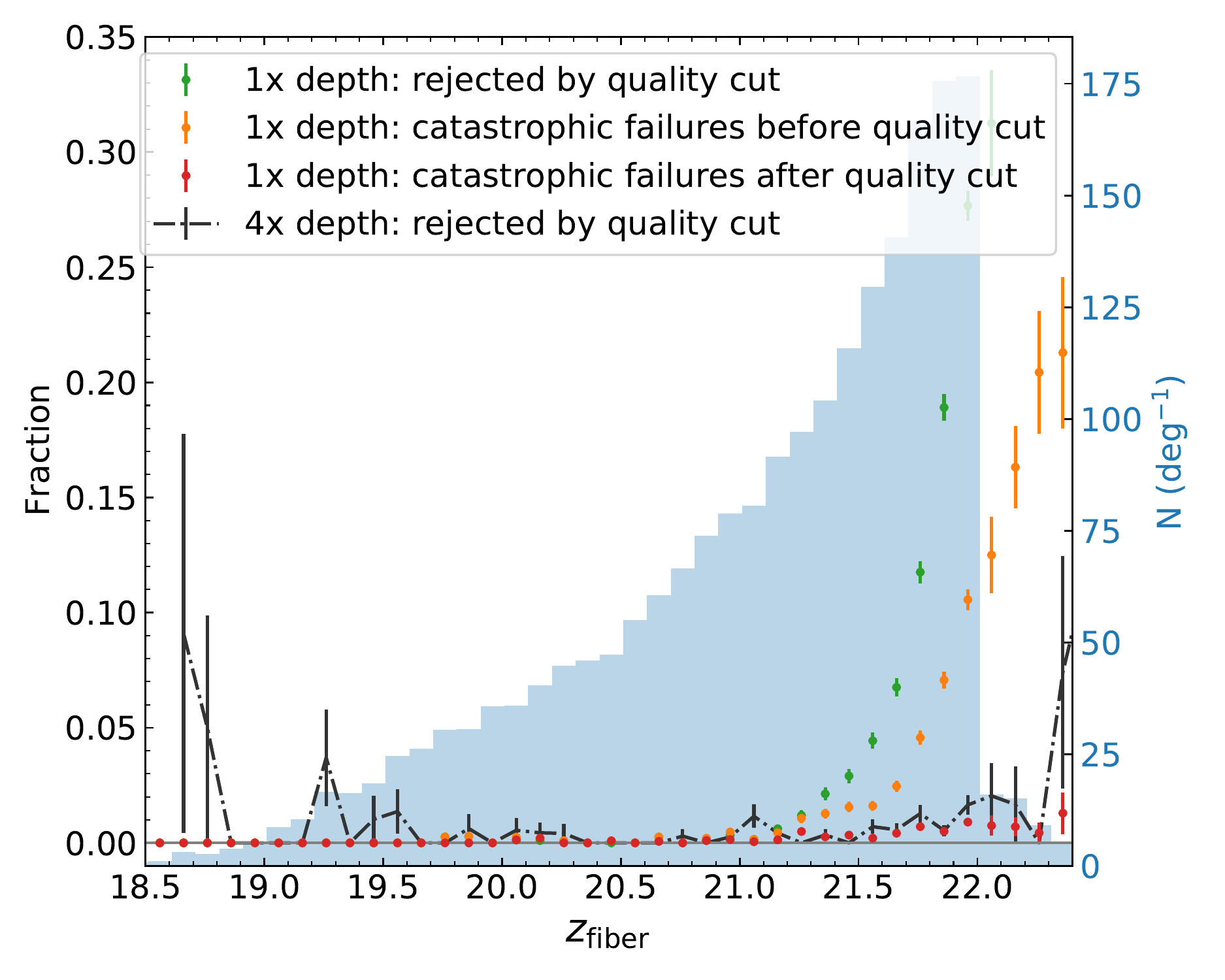}
    \caption{Catastrophic failure rates and rejection rates of SV1 LRGs as a function of $z$-band fiber magnitude for the nominal (1x-depth) exposures (points with error bars) and 4x-depth exposures (black dashed line with error bars). (Error bars are not shown for fractions equal to zero.) The histogram shows the $z_\mathrm{fiber}$ distribution of the SV1 LRG sample.}
    \label{fig:failure_rate_vs_zfiber_sv1}
\end{figure}

The major differences between the SV1 IR selection and the Main Survey selection are in the magnitude limit and the sliding color-magnitude cut ($r-W1$ vs $W1$). The SV1 magnitude limit is $0.4$ magnitude fainter than the Main sample, which increases the density at $z>{\sim}\,0.8$. The SV1 IR selection also has a more relaxed sliding color-magnitude cut that, combined with the fainter magnitude limit, more than doubles the comoving density of the Main sample. The redshift failure rate vs $z_\mathrm{fiber}$ of the SV1 LRGs is shown in Figure \ref{fig:failure_rate_vs_zfiber_sv1}. The failure/rejection rate increases drastically beyond $z_\mathrm{fiber}{\sim}\,21.6$ for the nominal DESI exposure time. The magnitude limit that we chose for the Main Survey is therefore a trade-off between redshift success completeness and number density at higher redshift.

We note that the failure rates as a function of $z_\mathrm{fiber}$ and $t_\mathrm{eff}$ are very similar to that of the Main sample, even though the comoving number density is much higher. This means that it is feasible to conduct a future survey with DESI (or a similar instrument) of a significantly denser sample of LRG-like galaxies at $z<{\sim}\,0.8$ without increasing the nominal exposure time. For instance, the ``SV1 bright'' sample, as shown in Figure \ref{fig:all_lrgs_dndz}, is a brighter subset of the SV1 LRGs that have the same magnitude limit as the Main LRGs and thus a very similar redshift success rate. The density of the ``SV1 bright'' sample is limited by the SV1 selection, and one can design a selection with a significantly higher density at $z<{\sim}\,0.8$). A higher comoving density at $z>0.8$ can be achieved with a fainter magnitude limit and longer exposure times.

SV1 includes magnitudes limits in both fiber magnitude and total magnitude, and this allowed us to compare the two faint limits. We find that, as expected, the fiber magnitude is a much better predictor of redshift success rate than the total magnitude, and a selection implementing a fiber magnitude limit includes significantly more high-z LRGs than is possible for a selection with similar redshift efficiency that implements a total magnitude limit. For this reason, we chose to set the faint limit in fiber magnitude for the Main Survey. As noted in \S\ref{sec:selection_cuts}, 
the fiber magnitude cut is effectively a morphology cut that could introduce systematics in both sample selection and theoretical modeling, and such effects should be carefully investigated in the clustering analysis.

\section{SV3 sample}
\label{sec:sv3}

Here we describe the LRG sample observed during the 1\% Survey (or ``SV3'') before the Main Survey operations. See \citet{dawson_sv_overview} for an overview of the 1\% Survey program.

The SV3 selection is designed after the decision is made to adopt the IR selection and before the Main selection is finalized. Therefore the SV3 selection is very similar to the Main selection but with small differences. The selection cuts are listed in Table \ref{tab:sv3_cuts} and shown in Figure \ref{fig:sv_lrg_selection}. The SV3 selection is based on the IR selection, as is the Main selection, although the sliding cut is slightly more extended. The magnitude limit is also fainter by ${\sim}\,0.1$ magnitude. As a result, the SV3 sample has a target density of ${\sim}\,800$ deg$^{-2}$, compared to the 605 deg$^{-2}$ of the Main sample. Although almost all Main LRGs are within the SV3 selection, ${\sim}\,2$ deg$^{-2}$ of main LRGs are not in SV3 because of differences in the low-z cuts (in $g-r$ vs $r-W1$, see panel (b) of Figure \ref{fig:sv_lrg_selection}). The SV3 program observed 140k unique SV3 LRG targets (excluding SV3 LRGs observed in BGS tiles).

\begin{table*}
    \centering
    \caption{SV3 selection cuts.}
    \label{tab:sv3_cuts}
    \begin{tabular}{ll}
    \hline
    \hline
    Cuts & Comment \\
    \hline
    \multicolumn{2}{c}{South}\\
    $z_\mathrm{fiber} < 21.7$ & Faint limit\\
    $z-W1 > 0.8\times(r-z) - 0.6$ & Stellar rejection\\
    $((g-r > 1.3) \ \mathrm{AND} \ (g-r > -1.55 \times (r-W1) + 3.13))$
    $\mathrm{OR} \ (r-W1 > 1.8)$ & Remove low-z galaxies\\
    $((r-W1 > (W1-17.26) \times 1.8)$
    $\mathrm{AND} \ (r-W1 > W1-16.36)) \ \mathrm{OR} \ (r-W1 > 3.29)$ & Luminosity cut\\
    \hline
    \multicolumn{2}{c}{North}\\
    $z_\mathrm{fiber} < 21.72$ & Faint limit\\
    $z-W1 > 0.8\times(r-z) - 0.6$ & Stellar rejection\\
    $((g-r > 1.34) \ \mathrm{AND} \ (g-r > -1.55 \times (r-W1) + 3.23))$
    $\mathrm{OR} \ (r-W1 > 1.8)$ & Remove low-z galaxies\\
    $((r-W1 > (W1-17.24) \times 1.83)$
    $\mathrm{AND} \ (r-W1 > W1-16.33)) \ \mathrm{OR} \ (r-W1 > 3.39)$ & Luminosity cut\\
    \hline
    \end{tabular}
\end{table*}

\section{Stellar Masses of DESI Legacy Imaging Survey Galaxies using Random Forests}\label{app:rf_masses}

To assess the stellar mass completeness of the LRG sample and guide our selection cuts, we
created a catalog of stellar masses of galaxies (not restricted to LRGs) in the DESI Legacy Imaging surveys. The catalog is general purpose and can be used for a variety of science cases, and it is publicly available\footnote{\url{https://data.desi.lbl.gov/public/ets/vac/stellar_mass/v1/}}.
In this section, we describe the data set and methods to estimate the stellar masses along with metrics used to assess their quality.

To estimate the stellar masses using only photometry from the DESI Legacy imaging surveys, we use a machine learning-based regression method called Random Forests \citep{Breiman2001RandomForest}. We train the model to take Milky Way extinction corrected colors $g-r$, $r-z$, $z-\mathrm{W}_{1}$, $\mathrm{W}_{1}-\mathrm{W}_{2}$ and the photometric redshifts from \citet{zhou_clustering_2021} as inputs to predict the stellar mass to observed-total-light ratio of a galaxy. The training set is comprised of galaxies in the Stripe 82 region whose photometry is available in the DESI Legacy Survey imaging catalog and stellar masses were measured by \citet{Bundy2015Stripe82Catalog} (S82-MGC) using SDSS $ugriz$ and UKIDS $YJHK$ photometry. We use the stellar mass estimates from S82-MGC, specifically the ``MASS\_OPT\_ZREIS'' value from the ``Mstar-z\_ukwide'' catalog\footnote{\url{https://www.ucolick.org/~kbundy/massivegalaxies/s82-mgc-catalogs.html}}, as the ``truth'' based on which we train the random forest model. Our mass estimates inherit any systematic uncertainty present in the stellar mass measurements from S82-MGC.

To decouple the effects of uncertainties in the photometric redshift from the prediction of stellar masses, we train our model to predict the stellar mass to observed-total-light ratio instead of the stellar mass. The photometric redshift is used to calculate a galaxy's luminosity distance which is then used to calculate the total light emitted by the galaxy in the $z$ band in the observer's frame of reference. The predictions of our model are then converted back to the stellar masses. To generate the catalog of stellar masses, we use only the objects in the DESI Legacy Survey imaging which have valid photometry (namely positive fluxes) in the $g$, $r$, $z$, $\mathrm{W}_{1}$, $\mathrm{W}_{2}$ bands and satisfy the stellar rejection cut of $r - W1 > 1.75 \times (r - z) - 1.1$. Objects that do not meet these requirements are assigned the value -99 in the stellar mass catalog.

To assess the accuracy of our stellar mass predictions, we divide the data set into 5 equal random subsets and train the model using 4 of those subsets combined and compare the predicted and true values of stellar masses for the remaining subset set. We repeat the same 5 times, each time selecting a different 4:1 split of the data. We calculate the performance metrics using the union of all 5 test sets. This process is also called 5-fold cross-validation. We quantify the accuracy of our predictions using the following three common metrics:

\begin{itemize}
    \item  \textbf{Prediction bias:} defined as $\langle\Delta \log_{10}(\mathrm{M}_{\star})\rangle$, where, $\Delta \log_{10}(\mathrm{M}_{\star}) = \log_{10}(\mathrm{True\ Stellar\ Mass})-\log_{10}(\mathrm{Predicted\ Stellar\ Mass})$. This quantifies the average error in our predictions.
    \item \textbf{Normalized Median Absolute Deviation ($\sigma_\mathrm{NMAD}$):} defined as $1.4826 \times \mathrm{Median}(\mid \Delta \log_{10}(\mathrm{M}_{\star}) \mid )$. $\sigma_\mathrm{NMAD}$ measure of the spread in our prediction which is also robust to outliers.
    \item \textbf{Fraction of Outliers (f$_{\mathrm{outlier}}$}:) defined as the fraction of stellar mass predictions for which $\mid \Delta \log_{10}(\mathrm{M}_{\star})\mid > 0.5\ dex$, i.e. the fraction of cases where the prediction error is very high. We chose the threshold of $0.05$ as it is a typical value of the systematic difference between many common stellar mass prediction methods.
 \end{itemize}

\begin{figure}
    \centering
    \includegraphics[width=0.9\columnwidth]{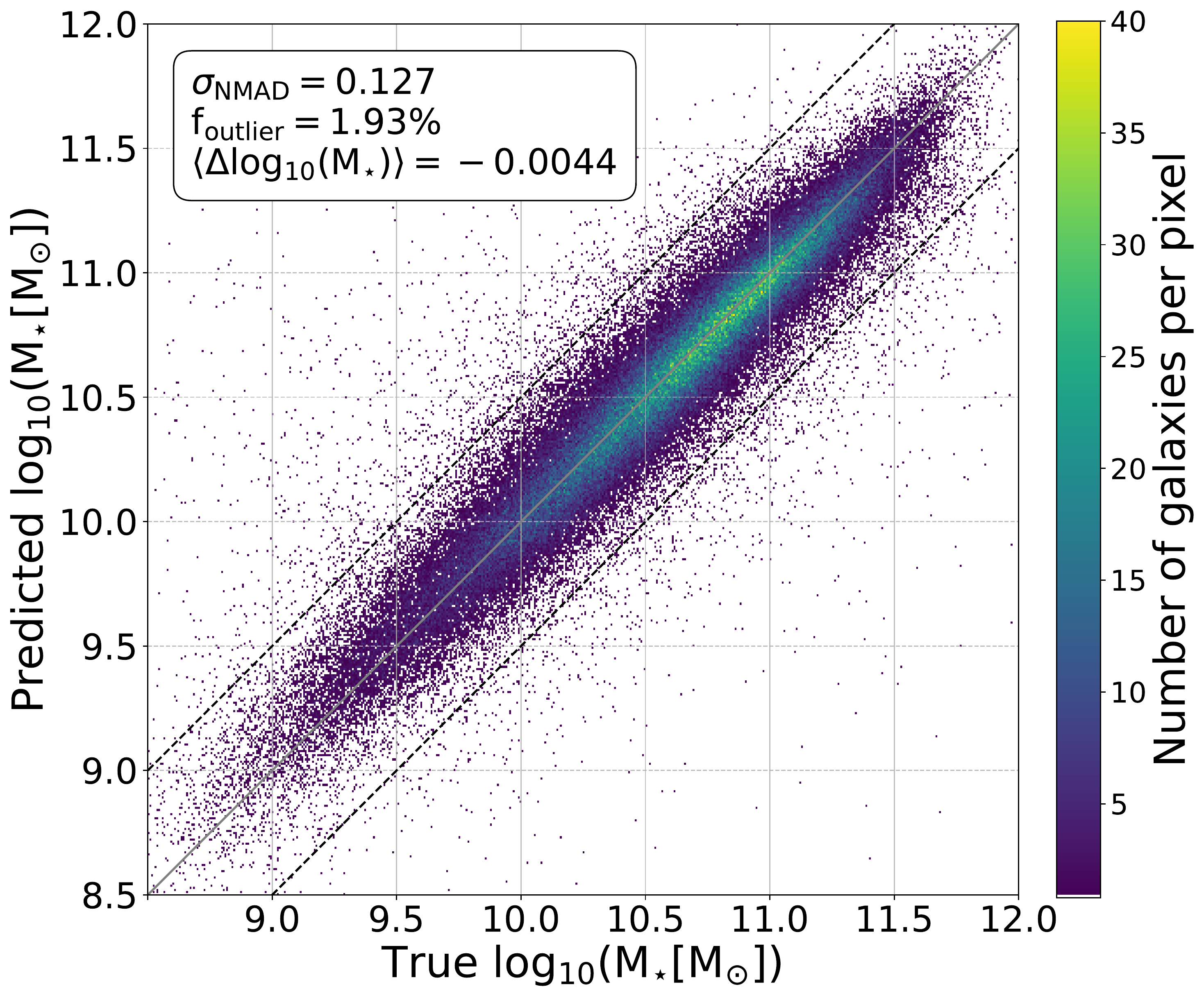}
    \caption{Comparison of predicted stellar mass and their true value (i.e. values from \citet{Bundy2015Stripe82Catalog}) for galaxies in the test set. The central gray line tracks the identity axis and the dashed lines mark the threshold beyond which a prediction is considered to be an outlier. The color of the points represents the number of data points present in each pixel of the figure. The scatter in the predictions is tight and symmetrically distributed about the identity line with a very small bias. The distribution of points is random and does not show any obvious pattern at the boundaries of training data indicating a stable performance across the entire range of stellar masses.}
    \label{fig:mass_comparison}
\end{figure}

\begin{figure}
    \centering
    \includegraphics[width=0.9\columnwidth]{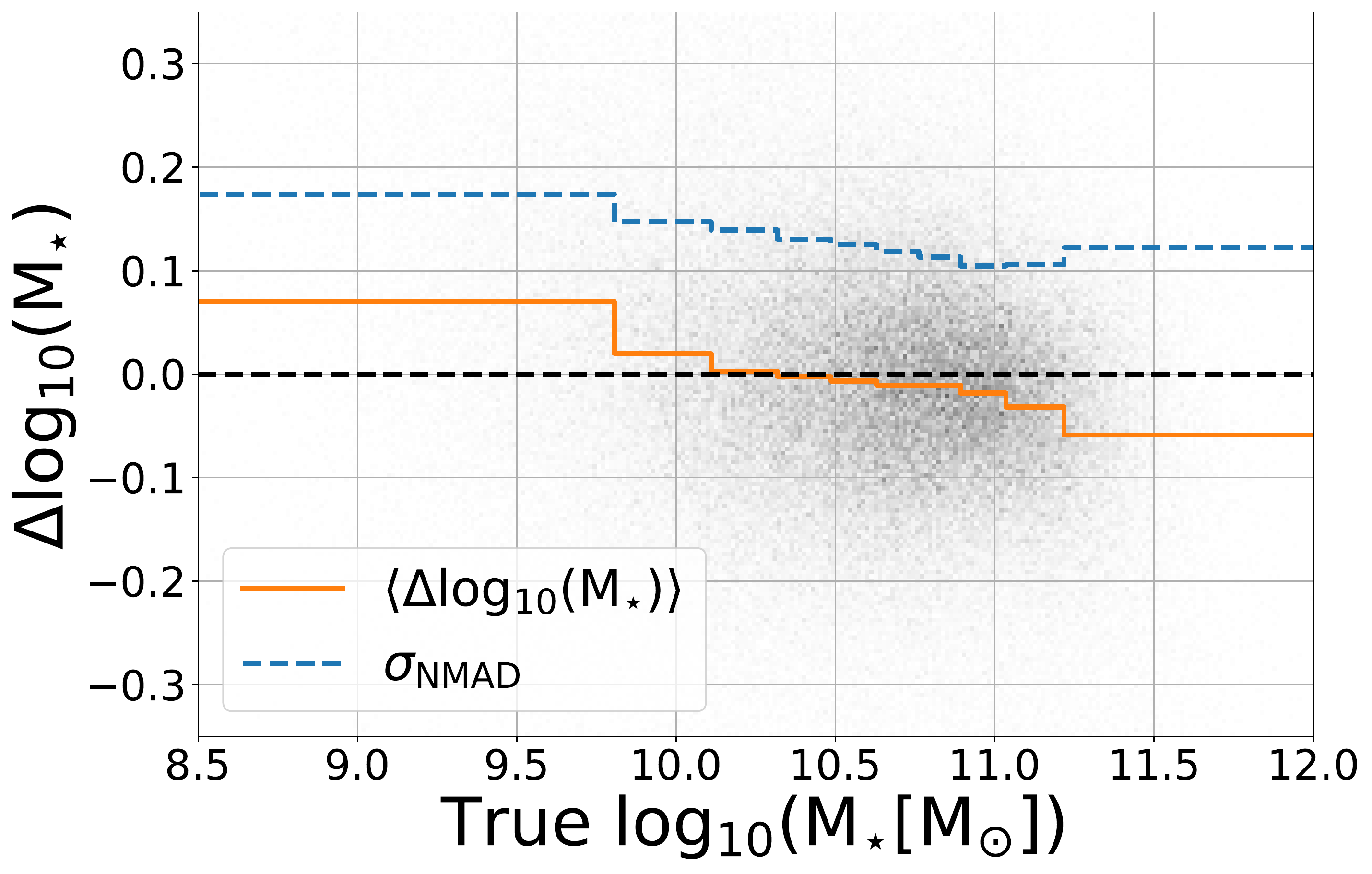}
    \caption{Evolution of prediction bias and $\sigma_{\mathrm{NMAD}}$ of our stellar mass predictions as a function of the true stellar masses. The metrics have been calculated for 10 equally populated bins with varying widths in stellar mass. This ensures that the standard errors on the binned statistics (even though very small due to a large number of data points per bin) are comparable across all bins. The gray scatter points show the distribution of galaxies in the test set. We observe that the magnitude of the prediction bias (i.e. $\langle\Delta \log_{10}(\mathrm{M}_{\star})\rangle$) bias stays under 0.1 dex and the scatter in predictions (quantified by $\sigma_{\mathrm{NMAD}}$) is between 0.2 and 0.1 dex for the entire range of stellar masses. These values of the metrics are typical of such comparisons between two methods of determining stellar mass from photometry (as shown in \citet{Bundy2015Stripe82Catalog}).}
    \label{fig:metric_v_mass}
\end{figure}

Our predictions have a moderate scatter of $\sigma_\mathrm{NMAD}=0.127$ dex, a small fraction of outliers (f$_{\mathrm{outlier}}=1.93\%$) with a negligibly small bias ($\langle\Delta \log_{10}(\mathrm{M}_{\star})\rangle=-0.0044$dex). As shown in Fig.~\ref{fig:mass_comparison}, the scatter in predictions are symmetric and does not show any obvious patterns at the boundaries of the training data. The prediction performance is also stable across the entire range of stellar masses (as shown in Fig.~\ref{fig:metric_v_mass}). We see that the prediction bias stays under 0.1 dex and $\sigma_{\mathrm{NMAD}}$ under 0.2 for the entire range of stellar masses. These values of the metrics are typical when two methods of estimating stellar masses are compared as shown in \citet{Bundy2015Stripe82Catalog}. 

If we instead train the model using the redshifts in S82-MGC and test using spectroscopic redshifts from DESI, we observe a significant improvement in the accuracy in stellar mass estimates with $\sigma_\mathrm{NMAD}=0.086$, f$_{\mathrm{outlier}}=0.75\%$ and $\langle\Delta \log_{10}(\mathrm{M}_{\star})\rangle=-0.0006$ dex. This indicates that the error in redshifts is a significant source of uncertainty for the stellar mass predictions. Since spectroscopic redshifts are only available for a subset of the objects in DESI Legacy Survey imaging, better quality mass predictions are available for only a fraction of objects. We notice that training with redshifts from S82-MGC and testing with photometric redshifts and vice versa results in poorer prediction accuracy and therefore the above combination of datasets was chosen to produce the two different mass catalogs. We use the catalog of stellar masses produced using photometric redshifts to analyze the stellar mass completeness of the DESI LRG sample.

\section{Veto masks for clustering analysis}
\label{sec:appdx_masks}

\begin{figure}
    \centering
    \includegraphics[width=0.9\columnwidth]{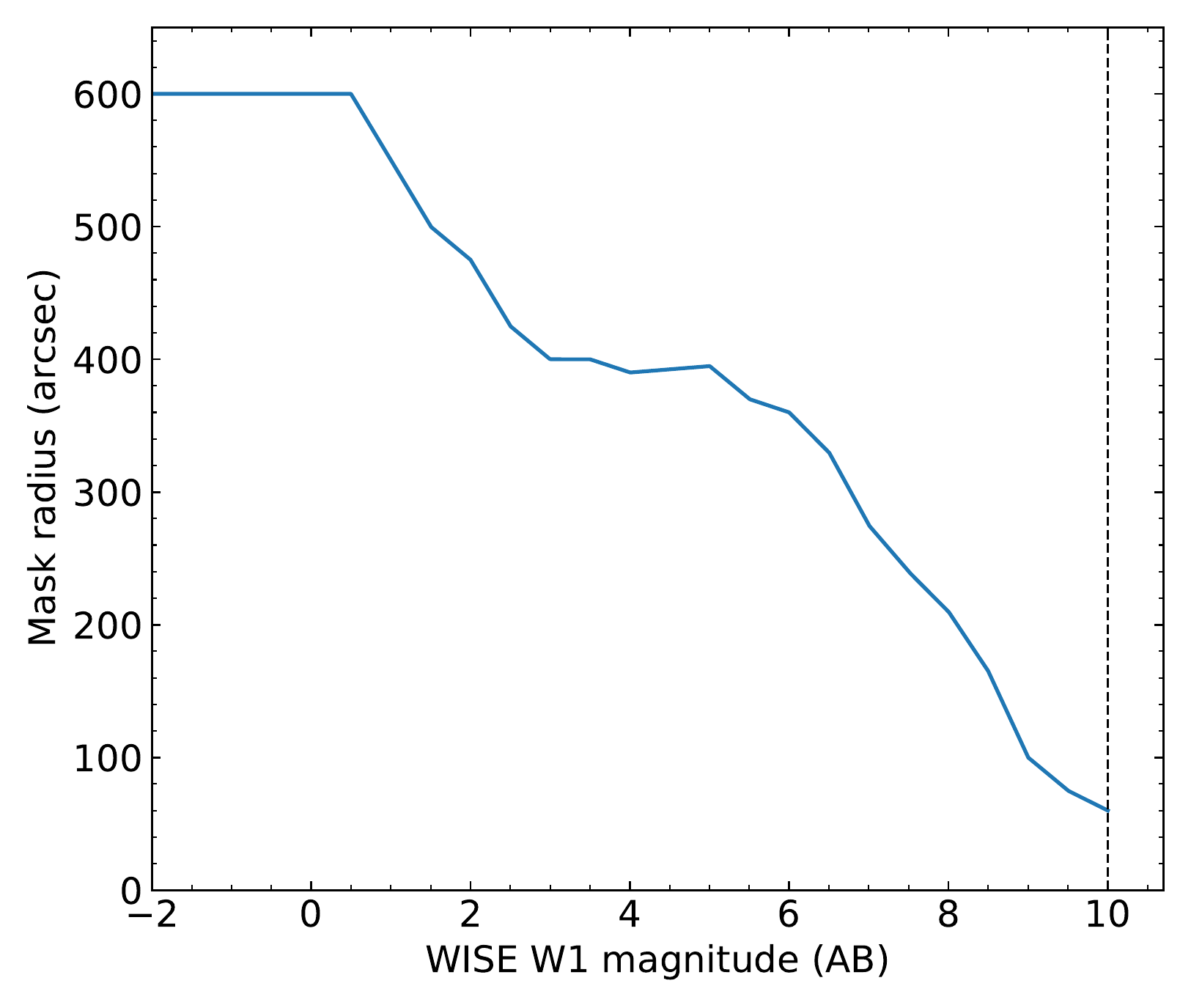}
    \includegraphics[width=0.9\columnwidth]{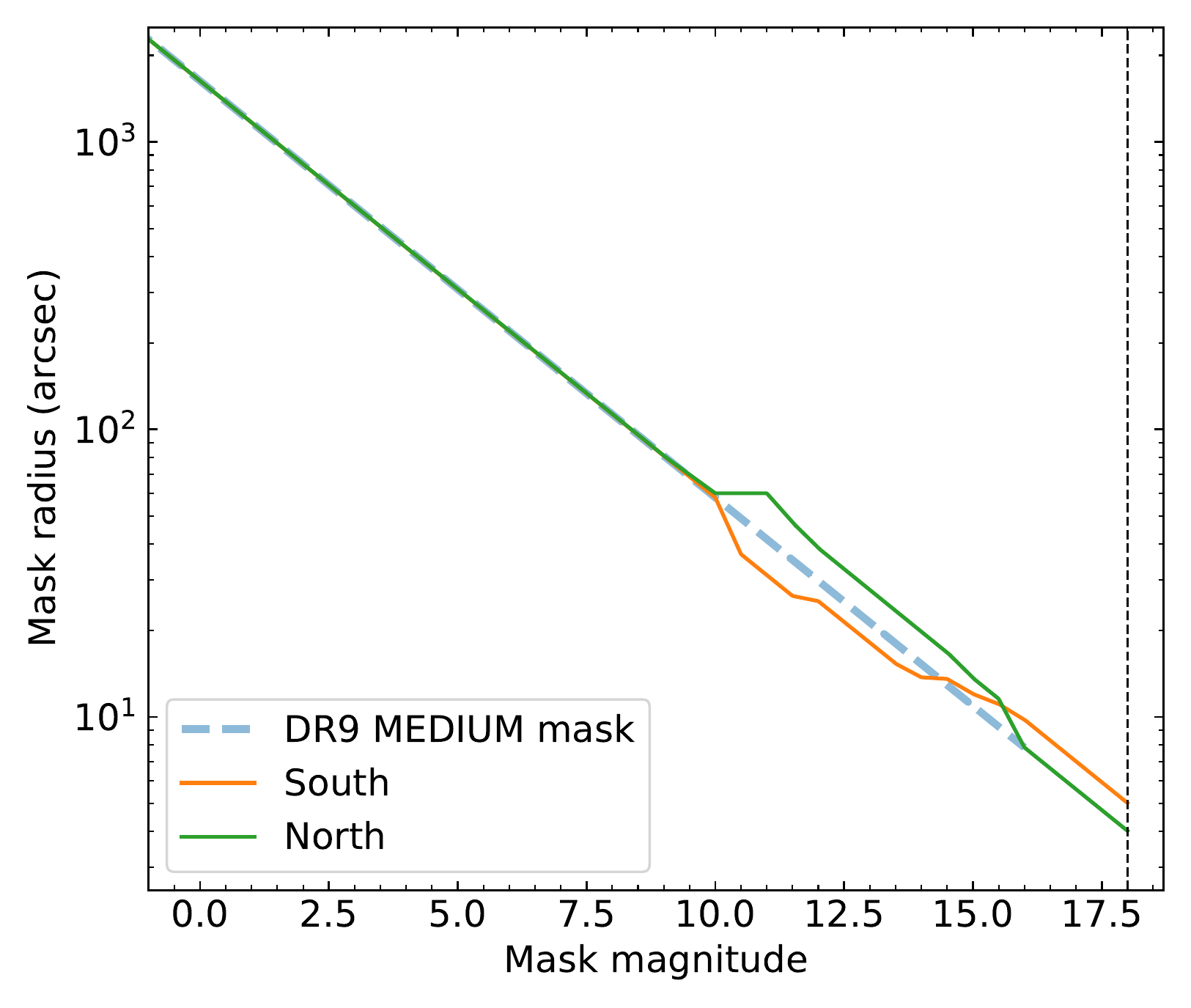}
    \caption{\textit{Top}: radius-magnitude relation for the WISE circular mask. \textit{Bottom}: radius-magnitude relation for the Gaia circular mask. The ``mask magnitude'' is typically Gaia $G$ magnitude, although for very red stars (with $z+1<G$ where $z$ is the predicted DECam $z$ magnitude), $z+1$ is used instead.) In both panels, The vertical dashed lines indicate the magnitude limit of the stars used in the masks.}
    \label{fig:mask_radius_vs_mag}
\end{figure}

The WISE circular geometric mask is based on the AllWISE star catalog (supplemented by 2MASS at the bright end) that was used for the unWISE mask \citep{meisner_unwise_2019}. Stars with a limiting magnitude of $W1<10.0$ are used. We derive the magnitude-radius relation empirically by checking the excess/deficit of LRG density around the stars in bins of magnitude and setting the mask radius to where the differential density excess/deficit is less than 10\%. In unWISE, the ``HALO'' mask radius depends not only on magnitude but also on ecliptic latitude and the sky background level. We find the trend of LRG densities around stars has no noticeable dependence on either of the two values, so we make the mask radius a function of magnitude only. The magnitude-radius relation for the WISE mask is shown in Figure \ref{fig:mask_radius_vs_mag} (top).

The Gaia mask is based on Gaia EDR3 with a limiting magnitude of $G<18$ (compared to $G<16$ in the LS DR9 MEDIUM mask) and supplemented with Tycho-2 stars at the bright end. Compared to Gaia DR2 (which was used in LS DR9), EDR3 has far fewer missing stars or stars with badly underestimated fluxes. And it contains few galaxies to $G<18$ (and those that are galaxies are at much lower redshift than the DESI LRGs), thus making the cut on astrometric excess noise (see \citealt{schlegel_dr9})
unnecessary. We use the Gaia $G$ magnitude as the mask magnitude (mask\_mag), except for very red stars that have zguess+1$<G$, where ``zguess'' is the predicted DECam $z$-band magnitude based on Gaia photometry (see \citealt{schlegel_dr9}). For these very red stars, we use zguess+1 as the mask magnitude, and this allows for sufficient masking for these stars.

Very few extremely bright stars are still missing from (or have incorrect photometry in) Gaia EDR3, and we supplement it with Tycho-2: if a Tycho-2 star with VT$<$10 is not within $1.0\arcsec$ of a Gaia star (with proper motion correction), we add it to the Gaia catalog. For these Tycho-2 stars, we predict the Gaia G magnitude (``ggguess'') and DECAm $z$ magnitude: we cross-match Tycho-2 to 2MASS to obtain the J-band photometry from 2MASS, derive the G-VT vs VT-J and z-VT vs VT-J relations (polynomials) using common stars in Tycho-2 and Gaia, and finally obtain ggguess and zguess for all Tycho-2 stars with 2MASS photometry. In the rare cases that a Tycho-2 star is not matched to 2MASS, the VT magnitude is used as the masking magnitude.

In LS DR9 the source detection and source fitting are slightly different inside the MEDIUM mask compared to outside the mask, and this could lead to slightly different LRG densities. Therefore, we also identify stars from the Gaia reference catalog in LS DR9 that have mask\_mag$<8$ and are brighter than the new Gaia catalog by more than 0.05 mag, and add them to the Gaia catalog, so that the mask radii of these stars are at least as large as their DR9 mask radii.

Similar to the WISE mask, we obtain the Gaia radius-magnitude relation empirically based on LRG densities around the stars. We see significant differences in LRG density trends around the stars between the North and the South (presumably due to differences in optics and detectors), in particular significantly more excess targets around stars of $10<G<16$ in the North, and we implement different radius-magnitude relations for the two regions. The magnitude-radius relations for the Gaia mask are shown in Figure \ref{fig:mask_radius_vs_mag} (bottom).

\begin{figure*}
    \centering
    \includegraphics[width=1.65\columnwidth]{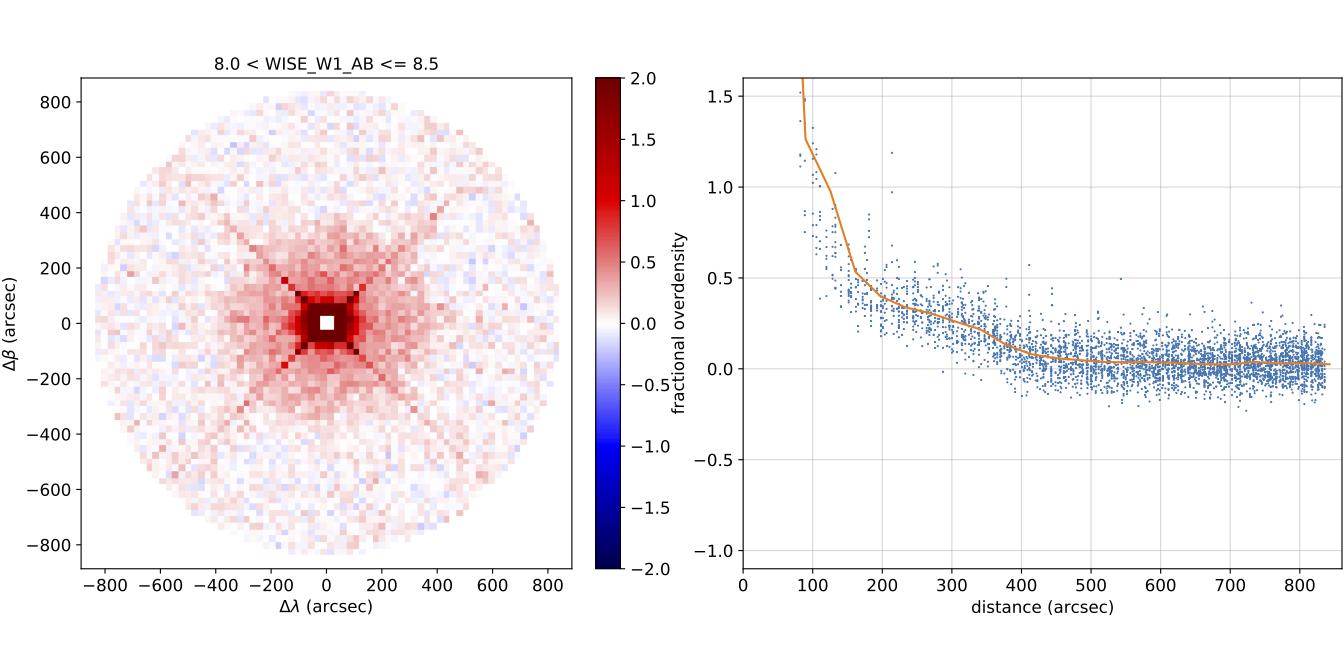}
    \includegraphics[width=1.65\columnwidth]{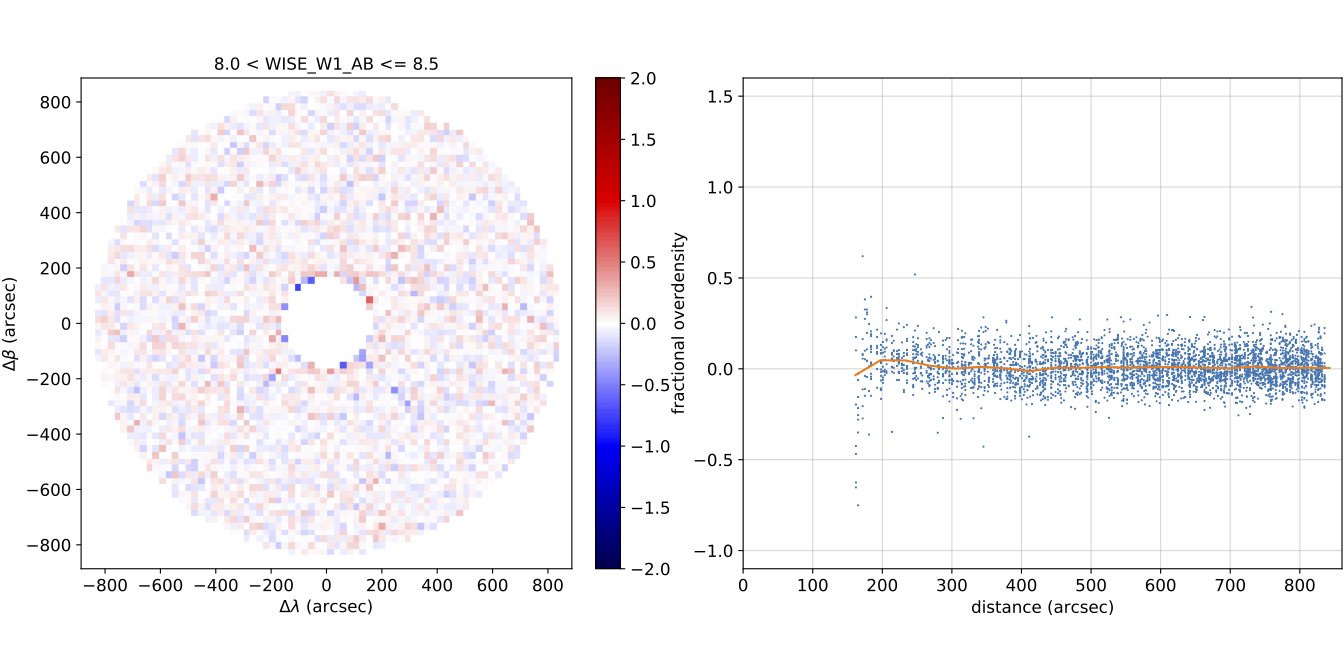}
    \caption{Relative LRG densities near WISE stars with $8<W1<8.5$ before (top panels) and after (bottom panels) applying the veto masks. The left panels show the fractional excess/deficit of LRG targets around the stars in 2-D bins of $\Delta\lambda$ and $\Delta\beta$ ($\lambda$ and $\beta$ are ecliptic coordinates). The right panels show the fractional excess/deficit as a function of distance for each $\Delta\lambda-\Delta\beta$ bin (dots) and their binned average (curve). The excess targets along the diffraction spikes are removed by the unWISE mask.}
    \label{fig:lrg_wise_cross_correlation}
\end{figure*}

\begin{figure*}
    \centering
    \includegraphics[width=1.65\columnwidth]{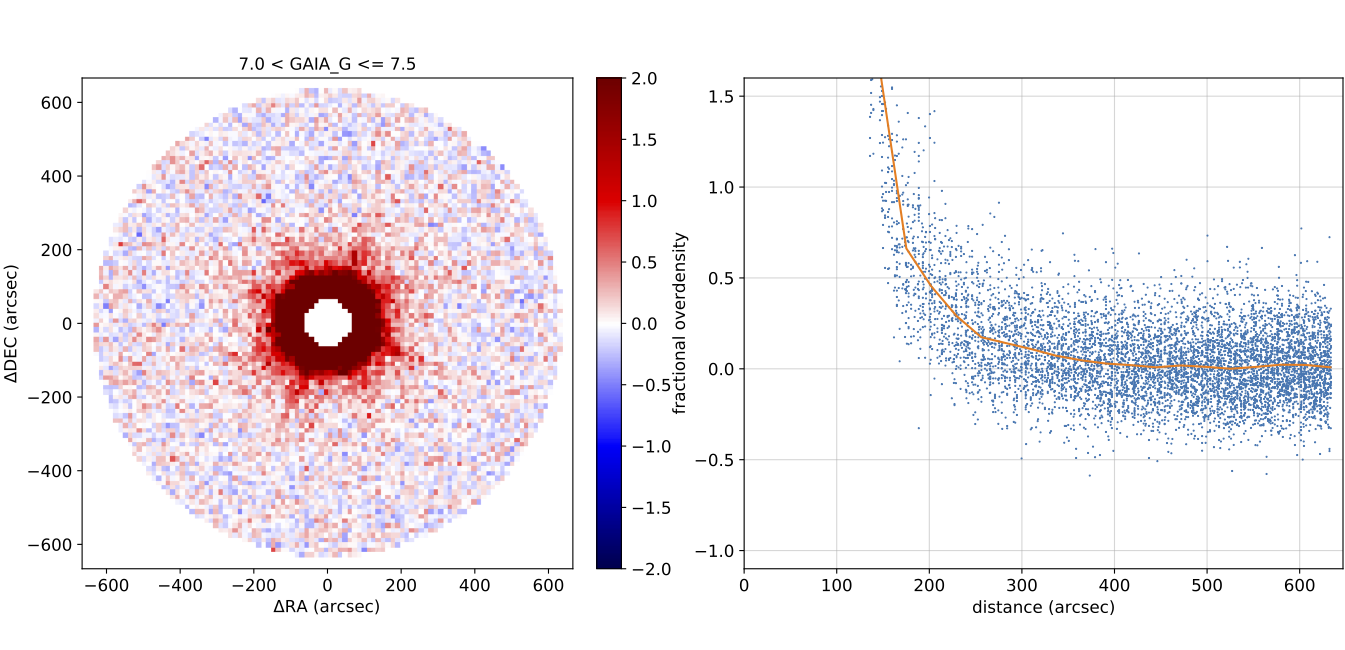}
    \includegraphics[width=1.65\columnwidth]{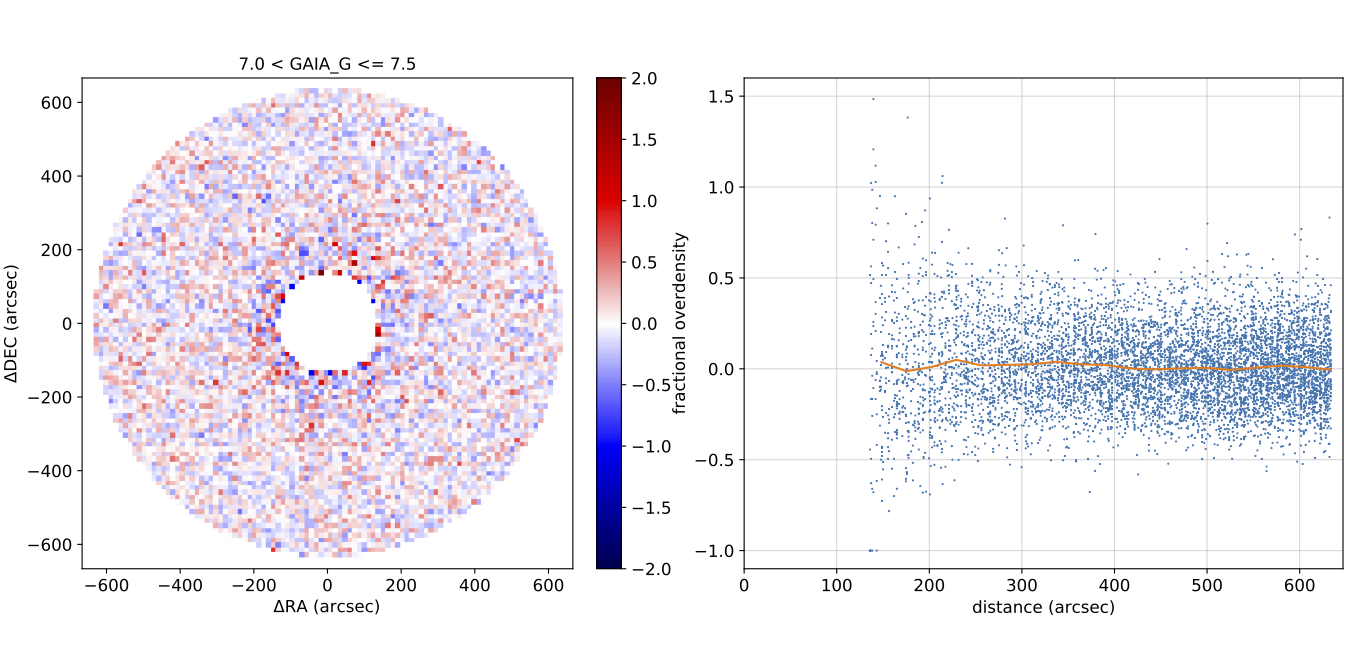}
    \caption{Same as Figure \ref{fig:lrg_wise_cross_correlation} but for Gaia stars with $7<G<7.5$ with the axes replaced by $\Delta$RA and $\Delta$DEC.}
    \label{fig:lrg_gaia_cross_correlation}
\end{figure*}

Figure \ref{fig:lrg_wise_cross_correlation} shows the LRG-WISE correlation for one of the WISE bins before and after applying the (full set of) veto masks. Figure \ref{fig:lrg_gaia_cross_correlation} shows the LRG-Gaia correlation for one of the Gaia bins before and after applying the (full set of) veto masks in the South.

In addition to the Gaia and WISE masks, we developed custom masks for any remaining problematic regions. Such regions include a small number of large galaxies that were not masked due to a bug in LS DR9 and regions with imaging artifacts that were found by visually inspecting regions identified by a DBSCAN cluster analysis \citep{Ester96adensity-based} as having very high LRG densities.

The stellar reference catalog (with mask radii) and the list of custom masks are publicly available\footnote{\url{https://data.desi.lbl.gov/public/ets/vac/lrg_veto_mask/v1/}}.

\bibliography{DESI_LRG_selection}{}
\bibliographystyle{aasjournal}

\end{document}